\documentclass[12pt]{article}
\pdfoutput=1

\usepackage{draft} 
\usepackage{hyperref}
\usepackage{graphicx,color,subfig}
\usepackage{cite}
\usepackage{mciteplus}
\usepackage{skak}
\usepackage{empheq}
\usepackage{tikz}
\usepackage{bbm}

\DeclareFontFamily{OT1}{pzc}{}
\DeclareFontShape{OT1}{pzc}{m}{it}{<-> s * [1.10] pzcmi7t}{}
\DeclareMathAlphabet{\mathpzc}{OT1}{pzc}{m}{it}

\usetikzlibrary{snakes}
\usetikzlibrary{arrows}
\usetikzlibrary{decorations.pathmorphing}
\tikzset{snake it/.style={decorate, decoration=snake}}
\usetikzlibrary{shapes.misc}
\tikzset{cross/.style={cross out, draw=black, minimum size=2*(#1-\pgflinewidth), inner sep=0pt, outer sep=0pt},
cross/.default={1pt}}

\def\be#1\ee{\begin{align}#1\end{align}}

\begin{document}

\unitlength = .8mm

\begin{titlepage}

\begin{center}

\hfill \\
\hfill \\
\vskip 1cm

\title{ZZ Instantons and the Non-Perturbative Dual \\ of $c=1$ String Theory}

\author{Bruno Balthazar, Victor A. Rodriguez, Xi Yin}

\address{
Jefferson Physical Laboratory, Harvard University, \\
Cambridge, MA 02138 USA
}

\email{bbalthazar@g.harvard.edu, victorrodriguez@g.harvard.edu, xiyin@fas.harvard.edu}

\end{center}

\abstract{ We study the effect of ZZ instantons in $c=1$ string theory, and demonstrate that they give rise to non-perturbative corrections to scattering amplitudes that do not saturate unitarity within the closed string sector. Beyond the leading non-perturbative order, logarithmic divergences are canceled between worldsheet diagrams of different topologies, due to the Fischler-Susskind-Polchinski mechanism. We propose that the closed string vacuum in $c=1$ string theory is non-perturbatively dual to a state of the matrix quantum mechanics in which all scattering states up to a given energy with no incoming flux from the ``other side" of the potential are occupied by free fermions. Under such a proposal, we find detailed agreement of non-perturbative corrections to closed string amplitudes in the worldsheet description and in the dual matrix model. }

\vfill

\end{titlepage}

\eject

\begingroup
\hypersetup{linkcolor=black}
\tableofcontents
\endgroup

\section{Introduction} 

The $c=1$ string theory \cite{Klebanov:1991qa, Ginsparg:1993is, Jevicki:1993qn, Polchinski:1994mb, Martinec:2004td} describes bosonic strings propagating in two spacetime dimensions, whose worldsheet theory consists of a free time-like boson $X^0$ and a $c=25$ Liouville theory, together with the $bc$ ghost system. Much of the interest in the subject has to do with the duality between $c=1$ string theory and a large $N$ gauged matrix quantum mechanics, which may be reformulated as the theory of a large number of free non-relativistic fermions subject to the Hamiltonian $H = {1\over 2} p^2-{1\over 2} x^2$, near a state in which the fermions fill the region $H<-\mu$, $x>0$ in the phase space. The closed strings are dual to collective excitations of the fermi surface. The agreement of perturbative scattering amplitudes computed in the worldsheet theory and in the matrix model has been explicitly demonstrated at tree level \cite{Moore:1991zv, DiFrancesco:1991ocm, DiFrancesco:1991daf, Balthazar:2017mxh}, and at 1-loop order \cite{Balthazar:2017mxh}. This duality has been extended to the gauge non-singlet sector of the matrix model \cite{Maldacena:2005hi, Balthazar:2018qdv}, whose states are mapped to ``long strings", defined as high energy limit of open strings attached to FZZT branes \cite{Fateev:2000ik, Teschner:2000md} that reside in the weak coupling region.

Conventionally, the above mentioned duality is thought to hold only at the level of perturbation theory, as the fermi sea is subject to the non-perturbative instability due to the fermion tunneling to the ``other side" ($x<0$) of the potential. It was proposed in \cite{McGreevy:2003kb} that a fermion near the top of the potential is dual to the (unstable) ZZ-brane \cite{Zamolodchikov:2001ah}. The open string tachyon excitation on the ZZ-brane allows it to access either side of the potential. This hints that string perturbation theory augmented by D-branes may give a non-perturbative formulation of the $c=1$ string.

In this paper, we find evidence that the worldsheet formulation of $c=1$ string theory can be completed non-perturbatively by taking into account D-instanton effects. Specifically, we consider ZZ-instantons, defined as a boundary condition on the worldsheet of Dirichlet type in $X^0$ direction and of ZZ type in the Liouville CFT, which are expected to be dual to the ``bounce" solution \cite{Coleman:1977py} describing the tunneling of a single fermion through the potential barrier. We will study the non-perturbative correction to the closed string $1\to n$ reflection amplitude mediated by a single ZZ-instanton, which renders the amplitude non-unitary. The leading contribution, of order $e^{-S_{\rm ZZ}}$, comes from $n+1$ disconnected discs, each emitting/absorbing a closed string, all of which have their boundaries ending on the same ZZ-instanton. In the $n=1$ case, we further analyze the next-to-leading order contribution, of order $e^{-S_{\rm ZZ}} g_s$, coming from (i) a single disc with two closed string insertions, and (ii) a disconnected diagram consisting of a disc and an annulus, each emitting/absorbing a closed string, such that divergences in the open string channel cancel between (i) and (ii) due to the Fischler-Susskind-Polchinski mechanism \cite{Polchinski:1994fq, Fischler:1986ci, Fischler:1986tb,Green:1997tv}.

\begin{figure}[h!]
\centering
\begin{tikzpicture}
\filldraw[color=black, fill=black!30, thick] (0,0) circle (1.25);
\draw (0,0) node[cross=4pt, very thick] {};
\draw (0,0) node[above] {$\cV^{+}_{\omega}$};
\draw[fill] (1.5,0) circle [radius=0.025];
\filldraw[color=black, fill=black!30, thick] (3,0) circle (1.25);
\draw (3,0) node[cross=4pt, very thick] {};
\draw (3,0) node[above] {$\cV^{-}_{\omega_1}$};
\draw[fill] (4.5,0) circle [radius=0.025];
\node at (5.25,0) {$\dots$};
\draw[fill] (6,0) circle [radius=0.025];
\filldraw[color=black, fill=black!30, thick] (7.5,0) circle (1.25);
\draw (7.5,0) node[cross=4pt, very thick] {};
\draw (7.5,0) node[above] {$\cV^{-}_{\omega_n}$};
\end{tikzpicture}
\caption{Worldsheet diagram that computes the leading non-perturbative correction to the closed string $1\to n$ amplitude, of order $e^{-S_{\rm ZZ}}$, mediated by a single ZZ-instanton.}
\label{fig:LO1ton}
\end{figure}
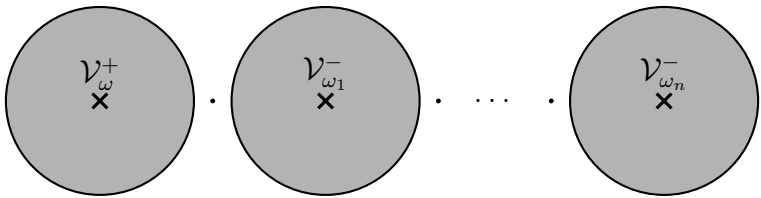

\begin{figure}[h!]
\centering
\begin{tikzpicture}
\filldraw[color=black, fill=black!30, thick] (0,0) circle (1.25);
\draw (-0.7,0) node[cross=4pt, very thick] {};
\draw (-0.7,0) node[above] {$\cV^{+}_{\omega_1}$};
\draw (0.7,0) node[cross=4pt, very thick] {};
\draw (0.7,0) node[above] {$\cV^{-}_{\omega_2}$};
\node at (2,0) {$+$};
\filldraw[color=black, fill=black!30, thick] (4,0) circle (1.25);
\draw (4,0) node[cross=4pt, very thick] {};
\draw (4,0) node[above] {$\cV^{+}_{\omega_1}$};
\draw[fill] (5.5,0) circle [radius=0.025];
\filldraw[color=black, fill=black!30, thick] (7,0) circle (1.25);
\filldraw[color=black, fill=white, thick] (7,0) circle (0.5);
\draw (6.2,0) node[cross=4pt, very thick] {};
\draw (6.2,0) node[above] {$\cV^{-}_{\omega_2}$};
\node at (9,0) {$+$};
\filldraw[color=black, fill=black!30, thick] (11,0) circle (1.25);
\filldraw[color=black, fill=white, thick] (11,0) circle (0.5);
\draw (10.2,0) node[cross=4pt, very thick] {};
\draw (10.2,0) node[above] {$\cV^{+}_{\omega_1}$};
\draw[fill] (12.5,0) circle [radius=0.025];
\filldraw[color=black, fill=black!30, thick] (14,0) circle (1.25);
\draw (14,0) node[cross=4pt, very thick] {};
\draw (14,0) node[above] {$\cV^{-}_{\omega_2}$};
\end{tikzpicture}
\caption{Worldsheet diagrams that compute the next-to-leading order non-perturbative correction to the closed string $1\to 1$ amplitude. The logarithmic divergences in the open string channel cancel between the connected and disconnected diagrams, as an example of the Fischler-Susskind-Polchinski mechanism.}
\label{fig:NLOdiagrams}
\end{figure}
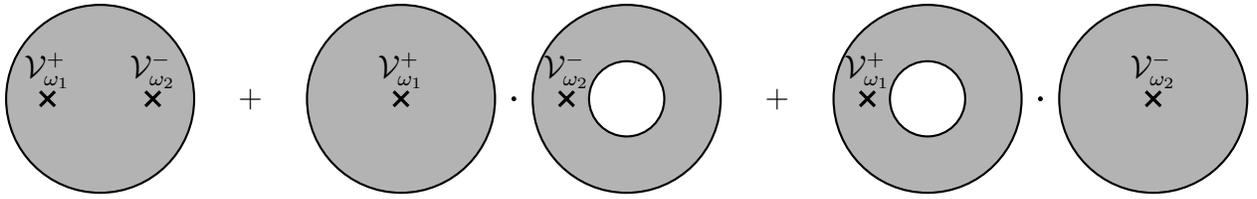

Our results for the leading one-instanton $1\to n$ amplitude and the next-to-leading order one-instanton $1\to 1$ amplitude agree, in a striking manner, with the non-perturbative correction to the reflection amplitude of a collective field in the dual matrix model, assuming that the fermi sea consists of only fermions that occupy scattering states with no incoming flux from the other side of the potential. We conjecture that the the latter is the correct dual of the closed string vacuum of the non-perturbative completion of $c=1$ string theory.

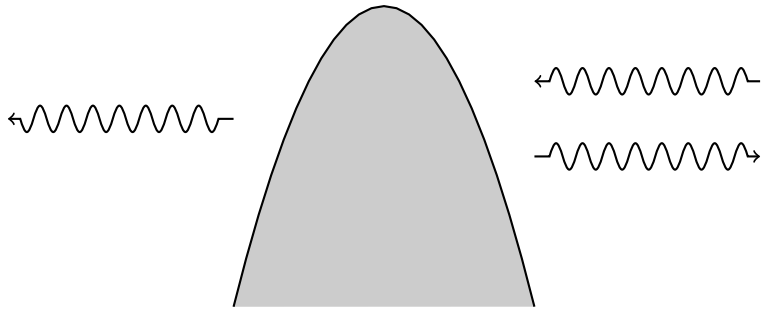
\begin{figure}[h!]
\centering
\begin{tikzpicture}
\draw[black, thick,fill=black!20, domain=-2:2] plot (\x, {-\x*\x});
\node at (0,1) {$ $};
\draw[<-, thick, snake=coil, segment aspect=0,segment amplitude=5pt, line before snake=2mm] (2,-1) -- (5,-1);
\draw[->, thick, snake=coil, segment aspect=0,segment amplitude=5pt, line before snake=2mm] (2,-2) -- (5,-2);
\draw[->, thick, snake=coil, segment aspect=0,segment amplitude=5pt, line before snake=2mm] (-2,-1.5) -- (-5,-1.5);
\end{tikzpicture}
\caption{The closed string vacuum is dual to the Fermi sea of scattering states with no incoming flux from the left side of the potential.}
\label{fig:clstrvac}
\end{figure}

The sense in which the ``non-perturbative corrections" are well-defined requires some explanation. While the perturbative closed string amplitudes of $c=1$ string theory obey perturbative unitarity, the ZZ-instanton effects generally renders the amplitude non-unitary within the closed string sector. The transition probability of closed string into other types of asymptotic states is unambiguously defined and is captured entirely by non-perturbative corrections. This is not the end of the story, however. The perturbative expansion of the closed $c=1$ string amplitude, as seen from the matrix model result, is an asymptotic series in $g_s$ that is Borel summable. According to our proposal, there are non-perturbative corrections to the Borel-resummed perturbative amplitude, and such corrections are captured by ZZ-instanton effects.


The unitary completion of the amplitude would require considering individual fermions on the other side of the potential among the asymptotic states, which are dual to rolling open string tachyons on the ZZ-brane \cite{Sen:2002nu, Sen:2004nf}. A detailed treatment of such states in the scattering problem is beyond the scope of this paper. Let us also emphasize that our proposal for the non-perturbative completion of $c=1$ string theory is {\it not} equivalent to the type 0B string theory/matrix model \cite{Takayanagi:2003sm, Douglas:2003up}. In the latter setting, the fermi sea fills in both sides of the potential, and the worldsheet theory involves an ${\cal N}=1$ superconformal field theory coupled to superconformal ghosts, subject to diagonal GSO projection, rather than the bosonic theory as considered here. 

This paper is organized as follows. In section \ref{zzsec} we outline the structure of ZZ-instanton contributions to closed string scattering amplitudes, and present explicit one-instanton computations at the leading and next-to-leading order in $g_s$. The results are then compared to the proposed matrix model dual in section \ref{mmsec}. After fixing certain finite counter terms (in the sense of open+closed string effective field theory on the ZZ-instanton), we find highly nontrivial agreement. Some future prospectives are discussed in section \ref{outlooksec}.

\section{Closed string amplitudes mediated by the ZZ instanton}
\label{zzsec}

The ZZ instanton is defined from the worldsheet perspective as a boundary condition that is of Neumann type in $bc$ ghosts, Dirichlet in the $X^0$ CFT, and ZZ boundary condition in the Liouville CFT. Its action $S_{\rm ZZ}$ is related to the mass $M_{\rm ZZ}= \mu = {1\over 2\pi g_s}$ of the ZZ-brane \cite{McGreevy:2003kb} (in $\A'=1$ units) by \cite{Polchinski:1995mt}
\ie
S_{\rm ZZ} = 2\pi M_{\rm ZZ} = {1\over g_s}.
\fe
This is equal to the action of the bounce solution \cite{Coleman:1977py} that computes the semi-classical tunneling probability $e^{-S_{\rm bounce}}$ of a fermion at the fermi surface $H=-\mu$ through the potential barrier,
\ie
S_{\rm bounce} = 2\int_{-\sqrt{2\mu}}^{\sqrt{2\mu}} dx \sqrt{2\mu - x^2} = 2\pi \mu.
\fe
We expect closed string amplitudes mediated by ZZ-instantons to be non-perturbative corrections that render the closed string S-matrix $S_c$ non-unitary due to tunneling effects. In particular, $1-S_c^\dagger S_c$ computes the probability of closed string states turning into other kinds of asymptotic states, such as ZZ-branes with rolling open string tachyon.

\subsection{The structure of D-instanton expansion and the Fischler-Susskind-Polchinki mechanism}
\label{sec:FSP}

The structure of the ZZ-instanton contribution to closed string amplitudes, as a special case of D-instanton amplitudes, was outlined in \cite{Polchinski:1994fq}. The one-instanton contribution to a closed string $n$-point connected amplitude, for instance, is computed by worldsheet diagrams with boundary components along which the string ends on the {\it same} ZZ-instanton, integrated over the collective coordinate of the ZZ-instanton. Note that, even though we are speaking of a contribution to the connected closed string S-matrix element, the worldsheet diagram need not be connected.

Let us first discuss a connected component of the worldsheet diagram with no closed string insertion. The ``empty" disc diagram with no closed string insertion represents $-S_{\rm ZZ}$. Formally, the summation over multiple disconnected discs, with symmetry factor taken into account, exponentiates to $e^{-S_{\rm ZZ}}$. With this understanding, we will simply drop all disconnected empty discs and include an overall factor of $e^{-S_{\rm ZZ}}$ in the one-instanton amplitude.

Connected components that are empty diagrams with non-negative Euler characteristic amount to quantum corrections (renormalization) of the ZZ-instanton action $S_{\rm ZZ}$, which are suppressed by powers of $g_s$. After taking this into account, the leading one-instanton contribution to the closed string amplitude comes from $n$ disconnected discs, each with a closed string insertion, of the form\footnote{This was already noted in the computation of D-instanton effects in type IIB superstring theory in \cite{Green:1997tv}.}
\ie
e^{-S_{\rm ZZ}} \int dx^0 \prod_{i=1}^n \left\langle {\cal V}_i \right\rangle^{D^2}_{{\rm ZZ},x^0},
\fe
where $\left\langle {\cal V}_i \right\rangle^{D^2}_{{\rm ZZ},x^0}$ is the disc amplitude of a single closed string vertex operator ${\cal V}_i$, with the appropriate ghost insertions, whose boundary condition is that of the ZZ-instanton. In particular, if ${\cal V}_i$ has energy $\omega_i$, e.g. ${\cal V}_i = g_s e^{i\omega_i X^0} V_{P_i = {|\omega_i|\over 2}}$ ($V_P$ being a primary of the Liouville CFT labeled by Liouville momentum $P$), then $\left\langle {\cal V}_i \right\rangle^{D^2}_{{\rm ZZ},x^0}$ depends on the collective coordinate $x^0$ through the factor $e^{i\omega_i x^0}$. The integration over $x^0$ then produces the delta function $\delta(\sum_i \omega_i)$ that imposes energy conservation. The resulting contribution to the $n$-point closed string amplitude is of order $e^{-{1/ g_s}}$, with no extra factors of $g_s$, as each disc comes with a factor of $1/g_s$ that cancels against the $g_s$ from the vertex operator.

At the next order in $g_s$, there will be contributions from the disc diagram with two closed string insertions, as well as the annulus diagram with one closed string insertion. For instance, there is a contribution to the $1\to1$ amplitude at order $e^{-{1\over g_s}} g_s$ of the form
\ie\label{eq:1to1}
e^{-S_{\rm ZZ}} \int dx^0 \left( \left\langle {\cal V}_1  {\cal V}_2 \right\rangle^{D^2}_{{\rm ZZ},x^0} + \left\langle {\cal V}_1 \right\rangle^{D^2}_{{\rm ZZ},x^0} \left\langle {\cal V}_2 \right\rangle^{A^2}_{{\rm ZZ},x^0} + \left\langle {\cal V}_2 \right\rangle^{D^2}_{{\rm ZZ},x^0} \left\langle {\cal V}_1 \right\rangle^{A^2}_{{\rm ZZ},x^0} \right),
\fe
where $\left\langle {\cal V}_i \right\rangle^{A^2}_{{\rm ZZ},x^0}$ is the annulus amplitude with a single closed string vertex operator ${\cal V}_i$, whose both boundaries end on the ZZ-instanton at time $x^0$.

\begin{figure}[h!]
\centering
\begin{tikzpicture}
\filldraw[color=black, fill=black!30, thick] (1.245,0.11) arc (5:355:1.25) -- (4-1.245,-0.11) arc (185:360:1.25) arc (0:175:1.25) -- (1.245,0.11);
\draw (0,0) node[cross=4pt, very thick] {};
\draw (0,0) node[above] {$\cV^{+}_{\omega_1}$};
\draw (4,0) node[cross=4pt, very thick] {};
\draw (4,0) node[above] {$\cV^{-}_{\omega_2}$};
\end{tikzpicture}
\medskip
\begin{tabular}{c c c}
& & \\
& & \\
\begin{tikzpicture}
\filldraw[color=black, fill=black!30, thick] (-3,0) circle (1.25);
\draw (-3,0) node[cross=4pt, very thick] {};
\draw (-3,0) node[above] {$\cV^{+}_{\omega_1}$};
\draw[fill] (-1.5,0) circle [radius=0.025];
\filldraw[color=black, fill=black!30, thick] (0,1) circle (1);
\filldraw[color=black, fill=white, thick] (0,1) circle (0.8);
\path [fill=black!30] (0,0) circle (1.25);
\draw [thick] (0.80,0.96) arc (50:180-50:1.25);
\draw [thick] (-0.96,0.80) arc (140:400:1.25);
\draw (0,0) node[cross=4pt, very thick] {};
\draw (0,0) node[above] {$\cV^{-}_{\omega_2}$};
\end{tikzpicture}
& $~~$ &
\begin{tikzpicture}
\filldraw[color=black, fill=black!30, thick] (0,1) circle (1);
\filldraw[color=black, fill=white, thick] (0,1) circle (0.8);
\path [fill=black!30] (0,0) circle (1.25);
\draw [thick] (0.80,0.96) arc (50:180-50:1.25);
\draw [thick] (-0.96,0.80) arc (140:400:1.25);
\draw (0,0) node[cross=4pt, very thick] {};
\draw (0,0) node[above] {$\cV^{+}_{\omega_1}$};
\draw[fill] (1.5,0) circle [radius=0.025];
\filldraw[color=black, fill=black!30, thick] (3,0) circle (1.25);
\draw (3,0) node[cross=4pt, very thick] {};
\draw (3,0) node[above] {$\cV^{-}_{\omega_2}$};
\end{tikzpicture}
\end{tabular}
\caption{Degeneration limit of the diagrams in Figure \ref{fig:NLOdiagrams} that leads to logarithmic divergences due to propagation of the open string collective mode on the ZZ-instanton. Such divergences cancel in the sum of the three diagrams, as an example of the Fischler-Susskind-Polchinski mechanism.}
\label{fig:divlimit}
\end{figure}
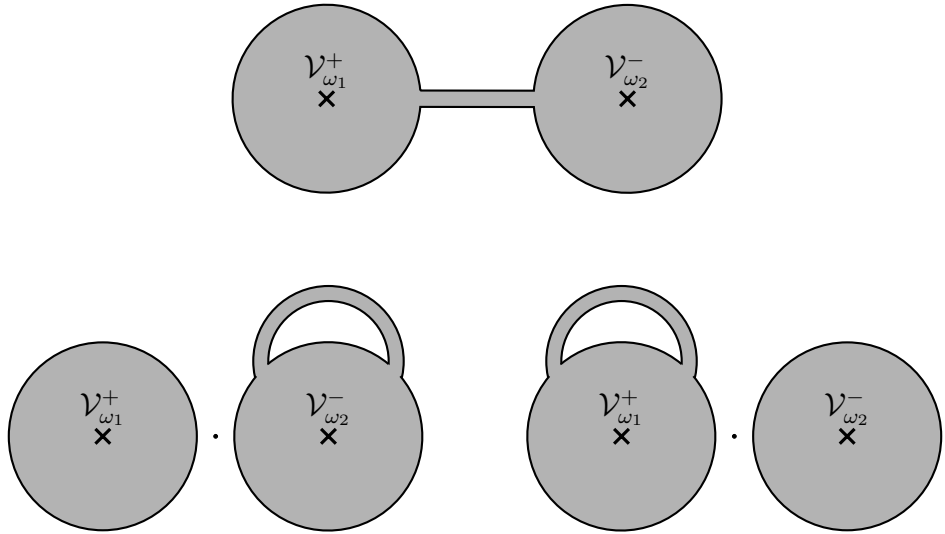

Each of the diagrams appearing in (\ref{eq:1to1}) suffers from divergence due to integration near the boundary of the moduli space, where a long/thin strip develops (Figure \ref{fig:divlimit}). There is a ``tachyonic" open string mode on the ZZ-instanton, whose propagation through the strip gives rise to a power divergence, which can be regularized by a simple subtraction as usual. On the other hand, the ``massless" open string mode on the ZZ-instanton, corresponding to the collective coordinate, gives rise to a logarithmic divergence. Such logarithmic divergences, as explained in \cite{Polchinski:1994fq, Green:1997tv}, cancel between diagrams of different topologies in (\ref{eq:1to1}).

In carrying out this cancelation mechanism, one must first regularize the logarithmic divergences by cutting off near the boundary of the moduli space, for each diagram in (\ref{eq:1to1}). There is a finite ambiguity in choosing the cutoff between diagrams of different topologies, proportional to
\ie\label{finiteambig}
& e^{-S_{\rm ZZ}} \int dx^0 \, \partial_{x^0} \left\langle {\cal V}_1 \right\rangle^{D^2}_{{\rm ZZ},x^0} \partial_{x^0} \left\langle {\cal V}_2 \right\rangle^{D^2}_{{\rm ZZ},x^0}.
\fe 
From the point of view of closed+open string field theory, where the open string fields live on the ZZ-instanton, this ambiguity is associated with a possible counter term in the open-closed string effective vertex that is compatible with the Batalin-Vilkovisky master equation. We do not know an a priori way to fix the constant coefficient of the ambiguity (\ref{finiteambig}) in the worldsheet computation of ZZ-instanton amplitudes. It will, however, be fixed through a comparison with the dual matrix model.

\subsection{Leading correction to the $1\to n$ closed string amplitude}

The $1\to n$ S-matrix element of $c=1$ string theory takes the form
\ie
S_{1\to n}(\omega; \omega_1, \cdots, \omega_n) = \delta\left(\omega-\sum_{i=1}^n\omega_i\right) {\cal A}_{1\to n}(\omega_1, \cdots, \omega_n),
\fe
where $\omega$ is the energy of the incoming closed string, and $\omega_1, \cdots, \omega_n$ are the energies of the reflected/outgoing closed strings. The unitarity relation takes the form
\ie
\int_{\omega_i\geq 0, ~\sum_{j=1}^n \omega_j = \omega} \prod_{i=1}^n d\omega_i {|{\cal A}_{1\to n}(\omega_1, \cdots, \omega_n)|^2\over \omega \omega_1\cdots \omega_n} = 1-P,
\fe
where $P$ is the probability of a closed string scattering into non-perturbative states. 

The leading correction to ${\cal A}_{1\to n}$ that violates unitarity within the closed string sector, thereby contributing to $P$, is due to the effect of a single ZZ-instanton, of order $e^{-{1/ g_s}}$. The relevant worldsheet diagram consists of $n+1$ disconnected discs, each with a closed string vertex operator insertion and with boundary ending on the same ZZ-instanton, as shown in Figure \ref{fig:LO1ton}.

The disc diagram with a single closed string vertex operator $\cV^\pm_\omega = g_s e^{\pm i \omega X^0} V_{\omega\over 2}$ insertion, corresponding to either an in-state or an out-state, is given by
\ie
\left\langle \cV^\pm_\omega(z,\bar z) \right\rangle^{D^2}_{{\rm ZZ},x^0}=g_s\frac{C_{D^2}}{2\pi}e^{\pm i\omega x^0} \Psi^{{\rm ZZ}}\left(\frac{\omega}{2}\right)=2e^{\pm i\omega x^0}\sinh(\pi\omega),
\label{eq:disk1pt}
\fe
where $x^0$ is the location of the ZZ-instanton in time, and $\Psi^{\text{ZZ}}$ is the disc 1-point function of the ZZ boundary state in Liouville CFT,
\ie
\Psi^{{\rm ZZ}}(P) = 2^{5\over 4} \sqrt{\pi} \sinh(2\pi P).
\fe
We have omitted writing the ghost insertions on the LHS. The factor of ${1\over 2\pi}$ after the first equality is due to division by the volume of the residual conformal Killing group. The normalization constant $C_{D^2}$ associated with the disc topology was determined in \cite{Balthazar:2018qdv}, and is given by
\ie
C_{D^2}=\frac{2^{\frac{3}{4}}\sqrt{\pi}}{g_s} .
\fe

The leading one-ZZ-instanton contribution to $S_{1\to n}(\omega; \omega_1,\cdots, \omega_n)$ is then evaluated as
\ie{}
S_{1\to n}^{{\rm inst},(0)} &= \cN \int_{-\infty}^{\infty} d x^0 e^{-S_{\rm ZZ}}\left\langle \cV^+_\omega(0) \right\rangle^{D^2}_{{\rm ZZ},x^0}\prod_{i=1}^n\left\langle \cV^-_{\omega_i}(0) \right\rangle^{D^2}_{{\rm ZZ},x^0}
\\
& = 2\pi \cN e^{-\frac{1}{g_s}}2^{n+1}\delta\left(\omega-\sum_{i=1}^n\omega_i\right)\sinh(\pi\omega)\prod_{i=1}^n \sinh(\pi\omega_i).
\label{eq:1toNws}
\fe
Note that there is an a priori unknown normalization factor $\cN$ in the integration measure over the collective coordinate $x_0$. In section \ref{mmsec}, this normalization factor will be fixed by comparison with the dual matrix model description\footnote{The normalization $\cN$ will be given by a real constant factor associated with the Lorentzian scattering amplitudes that we are computing. This is different from the Euclidean partition function of the ZZ-instanton, which receives an imaginary contribution from the open string tachyon.}.

\subsection{Next-to-leading order correction to the $1\to 1$ amplitude}
\label{sec:1to1NTLO}

We now analyze the next-to-leading order one-instanton correction to the closed string amplitude, of order $e^{-1/g_s} g_s$, focusing on the $1\to 1$ case. As discussed in (\ref{eq:1to1}), it receives contributions from the disc diagram with 2 closed string insertions, and from the disconnected diagram involving a disc and an annulus, each with 1 closed string insertion (Figure \ref{fig:NLOdiagrams}). Furthermore, there is a nontrivial cancelation of logarithmic divergences from individual diagrams (which must be carefully regularized).

\subsubsection{Disc 2-point diagram}
\label{disctwopt}

In the first diagram of Figure \ref{fig:NLOdiagrams}, we take the incoming closed string vertex operator $c\widetilde c {\cal V}_{\omega_1}^+$ to be fixed at $z=i$ on the upper half plane, whereas the outgoing closed string vertex operator ${c+\widetilde c\over 2} {\cal V}_{\omega_2}^-$ is fixed by the remaining conformal Killing vector to lie on the segment $y i$, $y\in(0,1)$.

The relevant free boson correlator on the upper half plane with Dirichlet boundary condition $\left.X^0\right|_{z=\bar z}=x^0$ is given by
\ie
 C^{X}_{D^2} 2^{\frac{\omega_1^2}{2}+\frac{\omega_2^2}{2}} y^{\frac{\omega_2^2}{2}}(1-y)^{\omega_1\omega_2}(1+y)^{-\omega_1\omega_2}e^{i (\omega_1-\omega_2)x^0},
\fe
while the ghost correlator is
\ie
2C^{bc}_{D^2}(1-y)(1+y).
\fe
Together with the Liouville correlator, the disc diagram gives (before integrating over the ZZ instanton collective coordinate)
\ie\label{vvxint}
& \left\langle {\cal V}^+_{\omega_1}{\cal V}^-_{\omega_2} \right\rangle^{D^2}_{{\rm ZZ},x^0}
\\
&=2g_s^2 C_{D^2}e^{i (\omega_1-\omega_2)x^0}2^{\frac{\omega_1^2}{2}+\frac{\omega_2^2}{2}}\int_0^1 dy \,  y^{\frac{\omega_2^2}{2}}(1-y)^{1+\omega_1\omega_2}(1+y)^{1-\omega_1\omega_2}\left\langle V_{\frac{\omega_1}{2}}(i)V_{\frac{\omega_2}{2}}(y i) \right\rangle^{D^2}_{\rm Liouville,ZZ}.
\fe
The Liouville correlator on the upper half plane with ZZ boundary condition can be evaluated by integrating the appropriate structure constants multiplied by Virasoro conformal blocks in either the ``bulk channel", where one first performs the OPE of the two bulk operators, or the ``boundary channel" where one first performs the bulk-to-boundary OPE. The relevant conformal blocks are simply specializations of sphere 4-point Virasoro conformal blocks. The bulk channel expression of the Liouville disc 2-point function is
\ie\label{bulkch}
\left\langle V_{\frac{\omega_1}{2}}(i)V_{\frac{\omega_2}{2}}(y i) \right\rangle^{D^2}_{\rm Liouville,ZZ}&=(1+y)^{-4-\omega_1^2}2^{\frac{\omega_1^2}{2}-\frac{\omega_2^2}{2}}y^{\frac{\omega_1^2}{2}-\frac{\omega_2^2}{2}}\int_0^\infty \frac{dP}{\pi}C\left(\frac{\omega_1}{2},\frac{\omega_2}{2},P\right)\Psi^{{\rm ZZ}}(P)
\\
&\times F\left(1+{\omega_1^2\over 4},1+{\omega_2^2\over 4},1+{\omega_1^2\over 4},1+ {\omega_2^2\over 4};1+P^2\left|\left(\frac{1-y}{1+y}\right)^2\right.\right),
\fe
where $C(P_1,P_2,P_3)$ is the DOZZ structure constant \cite{Dorn:1994xn, Zamolodchikov:1995aa} (following the convention of \cite{Balthazar:2017mxh, Balthazar:2018qdv}), and $F\left(h_1, h_2, h_3, h_4; h|z\right)$ is the $c=25$ sphere 4-point Virasoro conformal block in the $12\to 34$ channel with external primaries of weight $h_i$, $i=1,2,3,4$, internal primary of weight $h$, and cross ratio $z$.

The boundary channel only involves the identity operator as the internal boundary primary. The result is expressed as
\ie\label{bdrych}
&\left\langle V_{\frac{\omega_1}{2}}(i)V_{\frac{\omega_2}{2}}(y i) \right\rangle^{D^2}_{\rm Liouville,ZZ}=\\
&A(1+y)^{-4-\omega_1^2} 2^{\frac{\omega_1^2}{2}-\frac{\omega_2^2}{2}}y^{\frac{\omega_1^2}{2}-\frac{\omega_2^2}{2}}\Psi^{{\rm ZZ}}\left(\frac{\omega_1}{2}\right)\Psi^{{\rm ZZ}}\left(\frac{\omega_2}{2}\right)F\left(1+{\omega_1^2\over 4},1+{\omega_1^2\over 4},1+{\omega_2^2\over 4},1+{\omega_2^2\over 4};0\left|\frac{4y}{(1+y)^2}\right.\right),
\fe
where the numerical prefactor
\ie
\label{afac}
A=2^\frac{3}{4}\pi^{-{3\over 2}}
\fe
is fixed by crossing invariance between the bulk and boundary channels.\footnote{To derive (\ref{afac}), we can start from the crossing equation for the Liouville two-point function on the disk, multiply both sides by the leg-pole factor $S({\omega_1\over 2})^\frac{1}{2}$, where $S(P)\equiv- \left(\Gamma(2iP)/\Gamma(-2iP)\right)^2$ is the scattering phase off the Liouville wall, and then analytically continue $\omega_1\to2i$. In the latter limit, $S({\omega_1\over 2})^\frac{1}{2}C\left({\omega_1\over 2},{\omega_2\over 2},P\right)$ becomes proportional to $\delta\left({\omega_2\over 2}-P\right)$. The integral over $P$ in (\ref{bulkch}) is then easily evaluated to give the value of $A$ in (\ref{bdrych}).}

Let us analyze the limiting behavior of the $y$-integral in (\ref{vvxint}) near the boundary of the moduli space. The integral near $y=1$, using the bulk channel conformal block decomposition of the Liouville correlator (\ref{bulkch}), behaves as
\ie\label{xonelim}
& \int^1dy \int_0^\infty \frac{dP}{\pi}(1-y)^{-1+2P^2-\frac{(\omega_1-\omega_2)^2}{2}}C\left(\frac{\omega_1}{2},\frac{\omega_1}{2},P\right)\Psi^{{\rm ZZ}}(P)
\\
& \sim\int_0^\infty \frac{dP}{4P^2-(\omega_1-\omega_2)^2}C\left(\frac{\omega_1}{2},\frac{\omega_1}{2},P\right)\Psi^{{\rm ZZ}}(P).
\fe
Eventually we will integrate over the ZZ instanton collective coordinate $x^0$ (see section \ref{sec:coll}), which enforces $\omega_1=\omega_2$, in which case the $y$-integral converges near $y=1$. Indeed, $C(\frac{\omega_1}{2},\frac{\omega_1}{2},P)\Psi^{{\rm ZZ}}(P)\sim P^2$ for $P\sim0$, and so (\ref{xonelim}) converges near $P=0$; likewise one can verify that integration over large $P$ in (\ref{xonelim}) converges as well.

Near $y=0$, on the other hand, using the boundary channel conformal block decomposition of the Liouville correlator (\ref{bdrych}), one finds that (\ref{vvxint}) has a divergent part given by
\ie
g_s^2C_{D^2}\frac{A}{8}e^{i(\omega_1-\omega_2)x^0}\Psi^{{\rm ZZ}}\left(\frac{\omega_1}{2}\right)\Psi^{{\rm ZZ}}\left(\frac{\omega_2}{2}\right)\int_0 dy\, y^{-2}(1-2\omega_1\omega_2 y).
\label{eq:disk2div}
\fe
The leading power divergence is due to the open string ``tachyon" on the ZZ-instanton, and is simply subtracted in the usual regularization scheme. The subleading logarithmic divergence, coming from the open string collective mode (represented by the boundary vertex operator $\partial X^0$), cannot be removed by any local counter term. Rather, it will cancel against an analogous logarithmic divergence in the ``annulus $\times$ disc" diagrams in Figure \ref{fig:divlimit}, to be computed in section \ref{sec:anncomp}. 

For now, we simply regularize the divergence by the subtraction\footnote{Note that the subtraction of power divergence here is slightly non-standard: the subtraction involves the integral of $y^{-2}$ over the range $0<y<1$ rather than $0<y<\infty$. The difference between the two subtraction schemes can be absorbed into the finite ambiguity discussed in section \ref{sec:coll}, which will be fixed by comparison with the matrix model.}
\ie
\left\langle {\cal V}^+_{\omega_1}{\cal V}^-_{\omega_2} \right\rangle^{D^2, ~{\rm reg}}_{{\rm ZZ},x^0} 
&=2g_s^2 C_{D_2}e^{i (\omega_1-\omega_2)x^0}2^{\frac{\omega_1^2}{2}+\frac{\omega_2^2}{2}}
\\
& ~~~ \times\int_0^1 dy \left[ y^{\frac{\omega_2^2}{2}}(1-y)^{1+\omega_1\omega_2}(1+y)^{1-\omega_1\omega_2}\left\langle V_{\frac{\omega_1}{2}}(i)V_{\frac{\omega_2}{2}}(y i) \right\rangle^{D^2}_{\rm Liouville,ZZ} \right.
\\
&~~~~~~~~ \left.-\frac{A}{16}2^{-\frac{\omega_1^2+\omega_2^2}{2}}\Psi^{{\rm ZZ}}\left(\frac{\omega_1}{2}\right)\Psi^{{\rm ZZ}}\left(\frac{\omega_2}{2}\right)y^{-2}(1-2\omega_1\omega_2 y)\right].
\label{eq:disk2reg}
\fe
To evaluate (\ref{eq:disk2reg}), we use the boundary channel representation of the Liouville correlator (\ref{bdrych}) for $0<y<{1\over 2}$, and the bulk channel representation (\ref{bulkch}) for ${1\over 2}<y<1$, and perform the integration over the Liouville momentum $P$ and the modulus $y$ numerically. The details are discussed in Appendix \ref{sec:detailnumerics}.

\subsubsection{Annulus 1-point diagram}
\label{sec:anncomp}

The second and third diagrams of Figure \ref{fig:NLOdiagrams} are disconnected diagrams of a disc and an annulus with 1 closed string insertion on each. The disc 1-point diagram is calculated as in (\ref{eq:disk1pt}). We now turn to the annulus 1-point diagram.

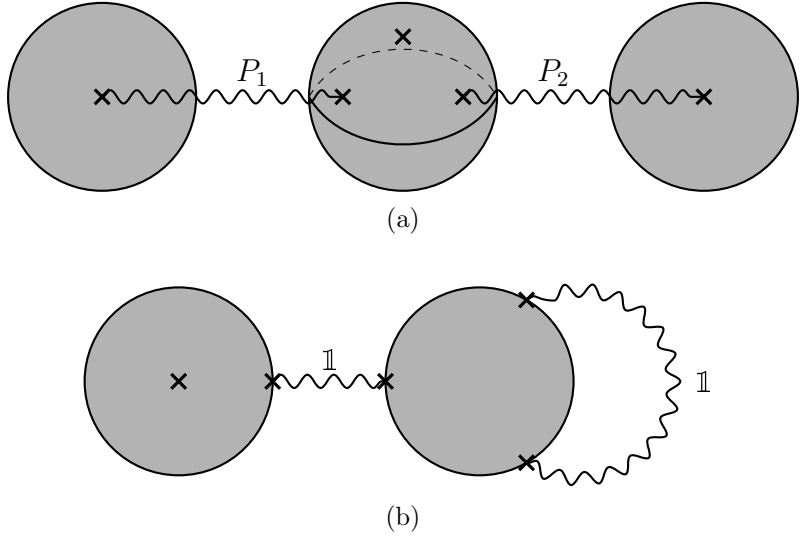
\begin{figure}[h!]
\centering
\subfloat[]{
\begin{tikzpicture}
\filldraw[color=black, fill=black!30, thick] (0,0) circle (1.25);
\draw[thick] (-1.25,0) to [out=-60,in=-120] (1.25,0);
\draw[dashed] (-1.25,0) to [out=60,in=120] (1.25,0);
\filldraw[color=black, fill=black!30, thick] (-4,0) circle (1.25);
\filldraw[color=black, fill=black!30, thick] (4,0) circle (1.25);
\draw (-4,0) node[cross=4pt, very thick] {};
\draw (-0.8,0) node[cross=4pt, very thick] {};
\draw (4,0) node[cross=4pt, very thick] {};
\draw (0.8,0) node[cross=4pt, very thick] {};
\draw (0,0.8) node[cross=4pt, very thick] {};
\draw[thick, snake it] (-4,0) -- (-0.8,0);
\draw[thick, snake it] (0.8,0) -- (4,0);
\draw (-2,0) node[above] {$P_1$};
\draw (2,0) node[above] {$P_2$};
\end{tikzpicture}
}\\
\medskip
\subfloat[]{
\begin{tikzpicture}
\filldraw[color=black, fill=black!30, thick] (0,0) circle (1.25);
\filldraw[color=black, fill=black!30, thick] (4,0) circle (1.25);
\draw (0,0) node[cross=4pt, very thick] {};
\draw (1.25,0) node[cross=4pt, very thick] {};
\draw (2.75,0) node[cross=4pt, very thick] {};
\draw (4+0.625,0+1.083) node[cross=4pt, very thick] {};
\draw (4+0.625,0-1.083) node[cross=4pt, very thick] {};
\draw[thick, snake it] (1.25,0) -- (2.75,0);
\draw[thick, snake it] (4+0.625,0-1.083) arc (240:480:1.25);
\draw (2,0) node[above] {$\mathbbm{1}$};
\draw (6.7,0) node[right] {$\mathbbm{1}$};
\end{tikzpicture}
}
\caption{Expansion of the Liouville bulk 1-point correlator on the annulus with ZZ boundary conditions in (a) the necklace channel, and (b) the OPE channel.}
\label{annuluschannels}
\end{figure}

We will parameterize the annulus as the strip $0<{\rm Re}(z)<\pi$, with the identification $z\sim z +2\pi i t $, $t>0$. We take the closed string vertex operator to be ${\cal V}_\omega^+$, and both boundary components of the annulus to end on the same ZZ instanton at $x^0$. The diagram is evaluated as
\ie
\left\langle {\cal V}^+_{\omega} \right\rangle^{A^2}_{{\rm ZZ},x^0} = g_s\int_0^\infty dt \int^{\frac{1}{4}}_0 2\pi dx \left\langle \frac{1}{4\pi}(b,\mu_t) \, c_2(2\pi x) : e^{i\omega X^0}: V_{P=\frac{\omega}{2}}(2\pi x)\right\rangle^{A^2(t)}_{{\rm ZZ},x^0}.
\fe
The vertex operator is fixed to be on the real axis, at $z=2\pi x$, accompanied by the ghost insertion $c_2 \equiv {c-\widetilde c\over 2i}$. We further used the symmetry $z\to \pi- z$ to restrict the $x$-integration range to $(0, {1\over 4})$. The $b$ ghost integrated against the Beltrami differential $\mu_t$ is given by
\ie
\frac{1}{4\pi}(b,\partial_t g)=-\pi(b+\tilde{b}).
\fe
The ghost correlator is evaluated similarly to that of the holomorphic ghost system on the torus via the doubling trick,
\ie\label{ghostcorr}
\left\langle \frac{1}{4\pi}(b,\mu_t) c_2(2\pi x) \right\rangle_{A^2(t)}=2\pi  C_{A^2}^{b,c}\eta(i t)^2,
\fe
where the overall phase is fixed by demanding that (\ref{ghostcorr}) is real and positive.
The $X^0$ part of the correlator evaluates to
\ie
\left\langle : e^{\pm i\omega X^0(2\pi x)}:\right\rangle_{A^2(t)}=
C_{A^2}^{X^0}\frac{e^{\pm i\omega x^0}}{\eta(it)}\left[\frac{2\pi}{\partial_\nu\theta_1(0|it)}\theta_1\left(2x|it\right)\right]^{\frac{\omega^2}{2}}.
\fe
Finally, the Liouville part of the annulus correlator can be expressed in terms of structure constants and Virasoro conformal blocks in either the necklace channel or the OPE channel (Figure \ref{annuluschannels}). By the doubling trick, the relevant conformal blocks are restrictions of the torus 2-point conformal blocks considered in \cite{Cho:2017oxl, Balthazar:2017mxh}.

In the necklace channel, we have
\ie\label{eq:Cylnecklace}
& \left\langle V_{P=\frac{\omega}{2}}(2\pi x)\right\rangle^{A^2(t)}_{\rm Liouviile,ZZ}
\\
& =t^{-2-\frac{\omega^2}{2}}\int_0^\infty \frac{dP_1}{\pi}\int_0^\infty \frac{dP_2}{\pi} C\left(\frac{\omega}{2},P_1,P_2\right)\Psi^{{\rm ZZ}}(P_1)\Psi^{{\rm ZZ}}(P_2)q_1^{h_1-\frac{c}{24}}q_2^{h_2-\frac{c}{24}}
\\
&~~~\times F^{\rm Necklace}\left(q_1,h_1=1+P_1^2,d_1=1+\frac{\omega^2}{4},q_2,h_2=1+P_2^2,d_2=1+\frac{\omega^2}{4}\right),
\fe
where $F^{\rm Necklace}(q_1, h_1, d_1, q_2, h_2, d_2)$ is the torus two-point $c=25$ Virasoro conformal block in the necklace channel, defined as in \cite{Cho:2017oxl} on a torus of modulus $\tau = i/t$ that is a 2-fold cover of the annulus. Here $q_1=e^{-\frac{2\pi}{t}+\frac{4\pi  x}{t}}$ and $q_2=e^{-\frac{4\pi  x}{t}}$ are a pair of plumbing parameters, $h_1, h_2$ are the internal primary weights, $d_1, d_2$ are the external primary weights. 

The OPE channel expression of the Liouville annulus 1-point function is given by a single Virasoro conformal block,
\ie{}
& \left\langle V_{P=\frac{\omega}{2}}(2\pi x)\right\rangle^{A^2(t)}_{\rm Liouviile,ZZ}
\\
&=A \Psi^{{\rm ZZ}}\left(\frac{\omega}{2}\right)q^{-\frac{c}{24}}\left(2\sin(2\pi x)\right)^{-2-\frac{\omega^2}{2}} F^{\rm OPE}\left(q,h_1=0,v,h_2=0,d_1=1+\frac{\omega^2}{4},d_2=1+\frac{\omega^2}{4}\right),
\label{eq:CylOPE}
\fe
where $F^{\rm OPE}(q, h_1, v, h_2, d_1, d_2)$ is the torus two-point OPE channel block defined in \cite{Cho:2017oxl}. Here the moduli are parameterized by $q=e^{-2\pi t}$ and $v=\frac{1-q_2}{q_2}$, with $q_2=e^{4\pi i  x }$. The numerical constant $A$ is the same as (\ref{afac}), as dictated by the crossing relation in the Liouville CFT with ZZ boundary condition. 



Putting these together, the annulus 1-point diagram is given by
\ie
\left\langle {\cal V}^+_{\omega} \right\rangle^{A^2}_{{\rm ZZ},x^0}= 4\pi^2 g_s C_{A^2}\int_0^\infty dt\, \eta(it)\int_0^{\frac{1}{4}}dx\left[\frac{2\pi}{\partial_\nu\theta_1(0|it)}\theta_1\left(2x|it\right)\right]^{\frac{\omega^2}{2}} e^{i\omega x^0} \langle V_{P=\frac{\omega}{2}}(2\pi x)\rangle^{A^2(t)}_{\rm Liouviile,ZZ}.
\label{eq:Cyl1}
\fe
where $\langle V_{P=\frac{\omega}{2}}(2\pi x)\rangle^{A^2(t)}_{\rm Liouviile,ZZ}$ can be evaluated using either (\ref{eq:Cylnecklace}) or (\ref{eq:CylOPE}).

Let us study the limiting behavior of the $t, x$ integral near the boundary of the moduli space. For $t$ away from $0$ and $\infty$, the integral near $x=0$ takes the form 
\ie
\pi^2 g_s C_{A^2}Ae^{i\omega x^0}\Psi^{{\rm ZZ}}\left(\frac{\omega}{2}\right)\int dt\int_0 dx\frac{e^{2\pi t}-1}{\sin^2(2\pi x)}\left[ 1+\mathcal{O}(x^2)\right].
\label{eq:Cylx0}
\fe
Here we see a power divergence associated to the open string tachyon mode on the ZZ instanton. We will regularize it by simply subtracting the term proportional to ${1\over \sin^2(2\pi x)}$ in the $x$-integrand. Possible finite ambiguities in this regularization scheme will be discussed in section \ref{sec:coll}. Note that there is no logarithmic divergence near $x=0$.

The integral near $t=\infty$ takes the form
\ie
\pi^2 g_s C_{A^2}Ae^{i\omega x^0}\Psi^{{\rm ZZ}}\left(\frac{\omega}{2}\right)\int^\infty dt\int_0^{1\over 4}dx \left[\frac{e^{2\pi t}-1}{\sin^2(2\pi x)}+2\omega^2\right].
\label{eq:tinf}
\fe
After subtracting off the power divergence in the $x$-integral near $x=0$, as discussed above, we end up with a subleading divergence of the form
\ie
\label{eq:AnnLog}
\frac{\pi^2}{2} g_s C_{A_2}A\omega^2 e^{i\omega x^0} \Psi^{{\rm ZZ}}\left(\frac{\omega}{2}\right)\int^\infty dt.
\fe
Comparing this limit to the disc 1-point amplitude also allows us to fix the normalization constant 
\ie
C_{A^2} = 1.
\fe
After multiplying by the disc 1-point diagram $\left\langle \cV^-_{\omega'}(z,\bar z) \right\rangle^{D^2}_{{\rm ZZ},x^0}$ (as given by (\ref{eq:disk1pt})), and integrating over the ZZ-instanton collective coordinate $x^0$, this divergence will cancel against the logarithmic divergence in the disc 2-point diagram (\ref{eq:disk2div}), provided that we make the identification
\ie\label{tyiden}
y \sim e^{-2\pi t}
\fe
near $y=0$ and $t=\infty$ respectively. In view of Figure \ref{fig:divlimit}, the degeneration limits of both the disc 2-point diagram and the disconnected disc$\times$annulus diagram can be viewed as a pair of discs attached to the same thin strip in different ways. (\ref{tyiden}) is then indeed the correct identification of moduli in the pinching limit, up to a finite constant ambiguity (multiplicative in $y$, or additive in $t$) that cannot be determined by the argument given so far.

Finally, let us consider integration near $t=0$. As discussed in appendix \ref{sec:Cylt0}, for real $\omega$ there are potential divergences due to the propagation of on-shell closed string states that need to be regularized. A simple way to treat this, as in \cite{Balthazar:2017mxh}, is to analytically continue from $\omega_i=\omega$ purely imaginary, where no such regularization is needed. Such an analytic continuation is straightforward in the domain $0<{\rm{Im}}(\omega)<1$ on the complex $\omega$-plane, where no poles cross the integration contour over internal Liouville momenta. One may then take the real $\omega$ limit to obtain the physical amplitude. 

For the sake of comparison with the matrix model dual, we will hereafter work with purely imaginary $\omega$ in the range $0<{\rm{Im}}(\omega)<1$, where the regularized annulus amplitude takes the form
\ie
\left\langle {\cal V}^+_{\omega} \right\rangle^{A^2}_{{\rm ZZ},x^0} &= 4\pi^2 g_s C_{A^2} e^{i\omega x^0} \int_0^\infty dt\int_0^{\frac{1}{4}}dx\Bigg\{ \eta(it) \left[\frac{2\pi}{\partial_\nu\theta_1(0|\tau)}\theta_1\left(2x|\tau\right)\right]^{\frac{\omega^2}{2}}  \left\langle V_{P=\frac{\omega}{2}}(2\pi x) \right\rangle^{A^2(t)}_{\rm Liouville,ZZ}
\\
&~~~~~~~~~~~~~~ -\frac{A}{4}\Psi^{{\rm ZZ}}\left(\frac{\omega}{2}\right)\left[ \frac{e^{2\pi t}-1}{\sin^2(2\pi x)}+2\omega^2\right] \Bigg\}.
\label{eq:Cyl1Imw}
\fe
The details of the numerical evaluation of (\ref{eq:Cyl1Imw}) are described in Appendix \ref{sec:detailnumerics}.

\subsubsection{Integration over collective coordinate and cancelation of divergences}
\label{sec:coll}

We can now put together the worldsheet diagrams in Figure \ref{fig:NLOdiagrams} and integrate over the ZZ-instanton collective coordinate $x^0$ as in (\ref{eq:1to1}) to obtain the contribution to $S_{1\to 1}$ at order $e^{-1/g_s} g_s$. As already discussed, the divergences in each diagram due to the exchange of open string collective mode cancel. It suffices to compute the finite part of the diagrams according to the regularized expressions (\ref{eq:disk2reg}) and (\ref{eq:Cyl1Imw}). There are, however, two possible further corrections at this order.


The first correction is due to the renormalization of $S_{\rm{ZZ}}$ in (\ref{eq:1to1}). The expression $e^{-S_{\rm ZZ}}$ with $S_{\rm{ZZ}}={1\over g_s}$ can be thought of as summing over empty disc diagrams. At subleading orders in $g_s$, there will be contributions from annulus diagrams of order $g_s^0$, followed by two-holed discs of order $g_s$, and so forth. Formally, the cylinder diagram renormalizes the measure factor ${\cal{N}}$ in the integration over the collective coordinate, e.g. in (\ref{eq:1toNws}). The two-holed disc effectively shifts $S_{\rm ZZ}$ by $g_s S_{\rm ZZ}^{(1)}$, so that the order $e^{-1/g_s} g_s$ contribution to $S_{1\to 1}$ is corrected by a term proportional to the ``disc 1-point squared" amplitude,
\ie\label{szzamb}
- e^{-1/g_s} g_s S_{\rm{ZZ}}^{(1)} 2\pi{\cal{N}}\delta(\omega_1-\omega_2)\left[\frac{g_sC_{D^2}}{2\pi}\Psi^{\rm{ZZ}}\left(\frac{\omega_1}{2}\right)\right]^2.
\fe
However, these diagrams without closed string insertions are divergent due to open string tachyons, and the regularization of the divergence is subject to finite ambiguities similar to the one encountered in the annulus 1-point diagram in section \ref{sec:anncomp}. Rather than trying to regularize all of these divergences simultaneously in a consistent manner, we will simply assume that $S_{\rm ZZ}^{(1)}$ in (\ref{szzamb}) is a constant, and will fix it along with ${\cal N}$ by comparing with the proposed matrix model dual.


The second correction to (\ref{eq:1to1}) has to do with the ambiguity in cutting off the logarithmic divergences due to the open string collective mode in the pinching limit of Figure \ref{fig:divlimit}. Namely, in identifying the cutoff parameters on the moduli $y$ versus $t$ according to (\ref{tyiden}), there is a finite constant ambiguity. This leads to an ambiguity in the order $e^{-1/g_s} g_s$ contribution to $S_{1\to 1}$ of the form
\ie\label{cpamb}
c' 2\pi {\cal{N}}\delta(\omega_1-\omega_2)g_s^2 C_{D^2}\frac{A}{4}\omega_1^2 \left[ \Psi^{{\rm ZZ}}\left(\frac{\omega_1}{2}\right) \right]^2.
\fe
We do not know of a worldsheet prescription that fixes $c'$ a priori. It will be determined by comparison with the matrix model dual as well.

Altogether, the next-to-leading order one-instanton correction to the $1\to 1$ closed strings amplitude is given by
\ie{}
& S_{1\to 1}^{{\rm inst},(1)}(\omega; \omega') 
\\
&= e^{-1/g_s} g_s \delta(\omega-\omega') 8\pi{\cal{N}} \Bigg\{ \pi^{1\over 2} 2^{-{1\over 4}+\omega^2 }\int_0^1 dy \bigg[ y^{\frac{\omega^2}{2}}(1-y)^{1+\omega^2}(1+y)^{1-\omega^2}\left\langle V_{\frac{\omega}{2}}(i)V_{\frac{\omega}{2}}(y i) \right\rangle^{D^2}_{\rm Liouville,ZZ} \\
&~~~~~~~~~~~~~~~~~~~~~~~~~~~~~~~~~~~~~~~~~~~~~~~~~~~~~~~~ -\frac{A}{16}2^{-\omega^2}\left( \Psi^{{\rm ZZ}}\left(\frac{\omega}{2}\right) \right)^2 y^{-2}(1-2\omega^2 y)\bigg ]\\
&~~~~~+2^{3\over 4}\pi^{3\over 2} \Psi^{\rm{ZZ}}\left(\frac{\omega}{2}\right)\int_0^\infty dt\int_0^{\frac{1}{4}}dx\Bigg[ \eta(it) \left(\frac{2\pi}{\partial_\nu\theta_1(0|\tau)}\theta_1\left(2x|\tau\right)\right)^{\frac{\omega^2}{2}} \left\langle V_{P=\frac{\omega}{2}}(2\pi x) \right\rangle^{A^2(t)}_{\rm Liouville,ZZ} \\
&~~~~~~~~~~~~~~~~~~~~~~~~~~~~~~~~~~~~~~~~~~~~~~ -\frac{A}{4}\Psi^{{\rm ZZ}}\left(\frac{\omega}{2}\right)\left( \frac{e^{2\pi t}-1}{\sin^2(2\pi x)}+2\omega^2\right) \Bigg] 
\\
&~~~~~ - S_{\rm ZZ}^{(1)} \sinh^2(\pi\omega) + c' \,\omega^2 \sinh^2(\pi\omega) \Bigg\}.
\label{eq:WSNLOall}
\fe

\section{Comparison with the matrix model dual}
\label{mmsec}

The matrix model dual to the closed string sector of $c=1$ string theory is defined as a suitable $N\to \infty$ scaling limit of the $U(N)$ gauged matrix quantum mechanics with Hamiltonian $H = {1\over 2} {\rm Tr}(P^2-X^2)$, where $X$ is a Hermitian $N\times N$ matrix and $P$ the canonically conjugate momentum matrix. Upon writing $X=\Omega^{-1}\Lambda\Omega$, where $\Omega\in U(N)$ and $\Lambda={\rm diag}(\lambda_1,\cdots, \lambda_N)$ a diagonal matrix, the wave function which is restricted to be $U(N)$-invariant can be expressed as a function of the eigenvalues $\Psi(\Lambda)$ that is completely symmetric in the $\lambda_i$'s. The Hamiltonian can be put to the form $H= \Delta^{-1} H' \Delta$, where $\Delta = \prod_{i<j} (\lambda_i - \lambda_j)$ and $H'= {1\over 2}\sum_i (-\partial_{\lambda_i}^2 - \lambda_i^2)$ is the Hamiltonian of $N$ free non-relativistic particles in the potential $V(\lambda) = -{1\over 2}\lambda^2$. As $H'$ effectively acts on the wave function $\Psi' = \Delta \Psi$, which is completely anti-symmetric in the $\lambda_i$'s, the eigenvalues are coordinates of free fermions.

Perturbatively, the closed string vacuum of $c=1$ string theory is dual to the state of the matrix model in which the fermions reside on the right side of the potential ($x>0$) and occupy all states up to the fermi energy $-\mu$, $\mu>0$. The string coupling is related by $2\pi g_s = \mu^{-1}$. Closed string states are dual to collective excitations of the fermi surface. Following the formalism of \cite{Moore:1991zv}, the collective modes can be represented as those of particle-hole pairs, and the S-matrix can be computed as a sum/integral over reflection amplitudes of individual particles and holes.

Non-perturbatively, the above prescription of the matrix model ground state is ill-defined. Rather than speaking of the fermions filling regions of the phase space, which only makes sense semi-classically, we will consider exact energy eigenstates of the fermion, which are scattering states of the Hamiltonian $H={1\over 2} p^2 - {1\over 2} x^2$. At a given energy $E$, there is a two-fold degeneracy: $|E\rangle_R$ which describes a scattering state with no incoming flux from the left side of the potential, and $|E\rangle_L$ which has no incoming flux from the right side. We propose that the non-pertubative closed string vacuum of $c=1$ string theory is dual to the state of the matrix model in which all scattering states of the form
\ie
|E\rangle_R,~~~E \leq - \mu,
\fe
and none other, are occupied by free fermions. Clearly, in the semi-classical limit (large $\mu$) this reduces to the picture of fermions filling the right fermi sea.

The exact reflection amplitude of a single fermion in frequency space is given by \cite{Moore:1991zv}\footnote{Note that $R(E)$ has poles on the lower half complex $E$-plane, as expected of S-matrix elements of non-relativistic scattering off a potential barrier.}
\ie
\label{exactre}
R(E) = i \mu^{iE} \left[ {1\over 1+e^{2\pi E}}\cdot {\Gamma({1\over 2}-i E)\over \Gamma({1\over 2} + i E)} \right]^{1\over 2}.
\fe
To compute the S-matrix of the collective modes, we also need to know the reflection amplitude of a hole. The ``hole scattering state" amounts to the un-occupation of the fermion scattering state $|E\rangle_R$. It describes the hole moving in the opposite direction of the would-be fermion in the state $|E\rangle_R$, with the roles of the incoming and outgoing waves exchanged. From this we deduce that the reflection amplitude of a hole is the inverse of that of a fermion, namely $(R(E))^{-1}$. While the fermion reflection amplitude $R(E)$ has magnitude less than 1 due to transmission/tunneling, the reflection amplitude of a hole has magnitude greater than 1. Note that this is different from the type 0B matrix model \cite{Takayanagi:2003sm, Douglas:2003up}, or the theory of ``type II" in \cite{Moore:1991zv}, where both sides of the fermi sea are filled and the reflection amplitude of a hole is $(R(E))^*$.

The exact $1\to n$ scattering amplitude is then given by \cite{Moore:1991zv}\footnote{In comparison to the analogous formulae in \cite{Moore:1991zv}, $R^*$ is replaced by $R^{-1}$ here.} 
\ie\label{ammoneton}
{\cal A}_{1\to n}(\omega_1,\cdots, \omega_n)= - \sum_{S_1\sqcup S_2=S}(-1)^{|S_2|}\int_0^{\omega(S_2)}dx \, R(-\mu+\omega-x)(R(-\mu-x))^{-1},
\fe
where $S_1,S_2$ are disjoint subsets of $S=\{\omega_1,...,\omega_n\}$ such that $S_1\sqcup S_2=S$. $|S_2|$ denotes the number of elements of $S_2$, and $\omega(S_2)$ the sum of all elements of $S_2$. We can write the integrand of (\ref{ammoneton}) as
\ie\label{rrfac}
R(-\mu+\omega-x)(R(-\mu-x))^{-1} = \left[ {1+ e^{-2\pi\mu} e^{-2\pi x} \over 1+ e^{-2\pi\mu} e^{2\pi(\omega-x)}} \right]^{1\over 2} K(\mu, \omega, x),
\fe
where
\ie
K(\mu, \omega, x) &\equiv \mu^{i\omega} \left[ {\Gamma({1\over 2} - i (-\mu+\omega-x)) \over \Gamma({1\over 2} - i (-\mu - x))} {\Gamma({1\over 2} + i (-\mu-x))\over \Gamma({1\over 2} + i(-\mu+\omega-x))} \right]^{1\over 2}.
\label{eq:Kfn}
\fe
The asymptotic series expansion of $K(\mu, \omega, x)$ in $\mu^{-1}$, integrated in $x$ according to (\ref{ammoneton}), gives the perturbative expansion of the matrix model amplitude that is expected to agree with the usual perturbative closed string amplitudes of $c=1$ string theory. Note that $K(\mu, \omega, x)$ is not analytic in $\mu$ at $\mu=\infty$. Nonetheless, one can verify that the ``perturbative part" of the amplitude, produced by the $x$-integral of $K(\mu, \omega, x)$, obeys the criteria of \cite{Sokal:1980ey}\footnote{This follows from the fact that $K(\mu, \omega, x)$ is analytic on the open domain ${\rm Re}\,\mu > \Lambda$ for some sufficiently large positive $\Lambda$ (at fixed $\omega$ and $x$), approaching 1 as $\mu\to\infty$, and that the error of the asymptotic series of $\log K$ in $\mu^{-1}$ truncated to order $\mu^{-N}$ in this domain is uniformly bounded by $C N! (\sigma/\mu)^N$ for some positive constants $C$ and $\sigma$, as can seen from the integral representation of log-Gamma function.} and is thus given by the Borel summation of the perturbative series by a theorem of Nevanlinna \cite{Nevanlinna}. 

The non-perturbative corrections comes from the prefactor on the RHS of (\ref{rrfac}). We can expand (\ref{rrfac}) as
\ie\label{rrexpa}
& R(-\mu+\omega-x)(R(-\mu-x))^{-1} \\
& = K(\mu, \omega, x) - e^{-2\pi \mu} e^{\pi (\omega - 2x)} \sinh(\pi\omega) \left[ 1 - {i\omega(x-{\omega\over 2})\over \mu} +{\cal O}(\mu^{-2}) \right] + {\cal O}(e^{-4\pi\mu}).
\fe
The term of order $e^{-2\pi \mu} = e^{-1/g_s}$ gives rise to
the leading non-perturbative contribution to the $1\to n$ amplitude
\ie
{\cal A}_{1\to n}^{{\rm inst}, (0)}(\omega_1, \cdots , \omega_n)= - 2^{n+1}\frac{e^{-2\pi\mu}}{4\pi}\sinh(\pi\omega)\prod_{i=1}^n\sinh(\pi\omega_i).
\label{eq:1toNmm}
\fe
This is in precise agreement with the worldsheet amplitude of disconnected discs with boundaries ending on a single ZZ-instanton (\ref{eq:1toNws}), up to the normalization constant ${\cal N}$ in the integration measure of the ZZ-instanton collective coordinate. Identifying the two results determines 
\ie
\cN= - \frac{1}{8\pi^2}.
\label{eq:N}
\fe
The term of order $e^{-2\pi\mu} \mu$ on the RHS of (\ref{rrexpa}), upon integration in $x$, gives the following next-to-leading order non-perturbative contribution to the $1\to 1$ amplitude
\ie
{\cal A}_{1\to 1}^{{\rm inst}, (1)}(\omega)= -i \frac{e^{-2\pi\mu}}{2\pi^2\mu}\omega\left(\frac{\pi\omega}{\tanh(\pi\omega)} - 1\right)\sinh^2(\pi\omega).
\label{eq:MMnlo}
\fe

\begin{figure}[h!]
\centering
\includegraphics[width=0.7\textwidth]{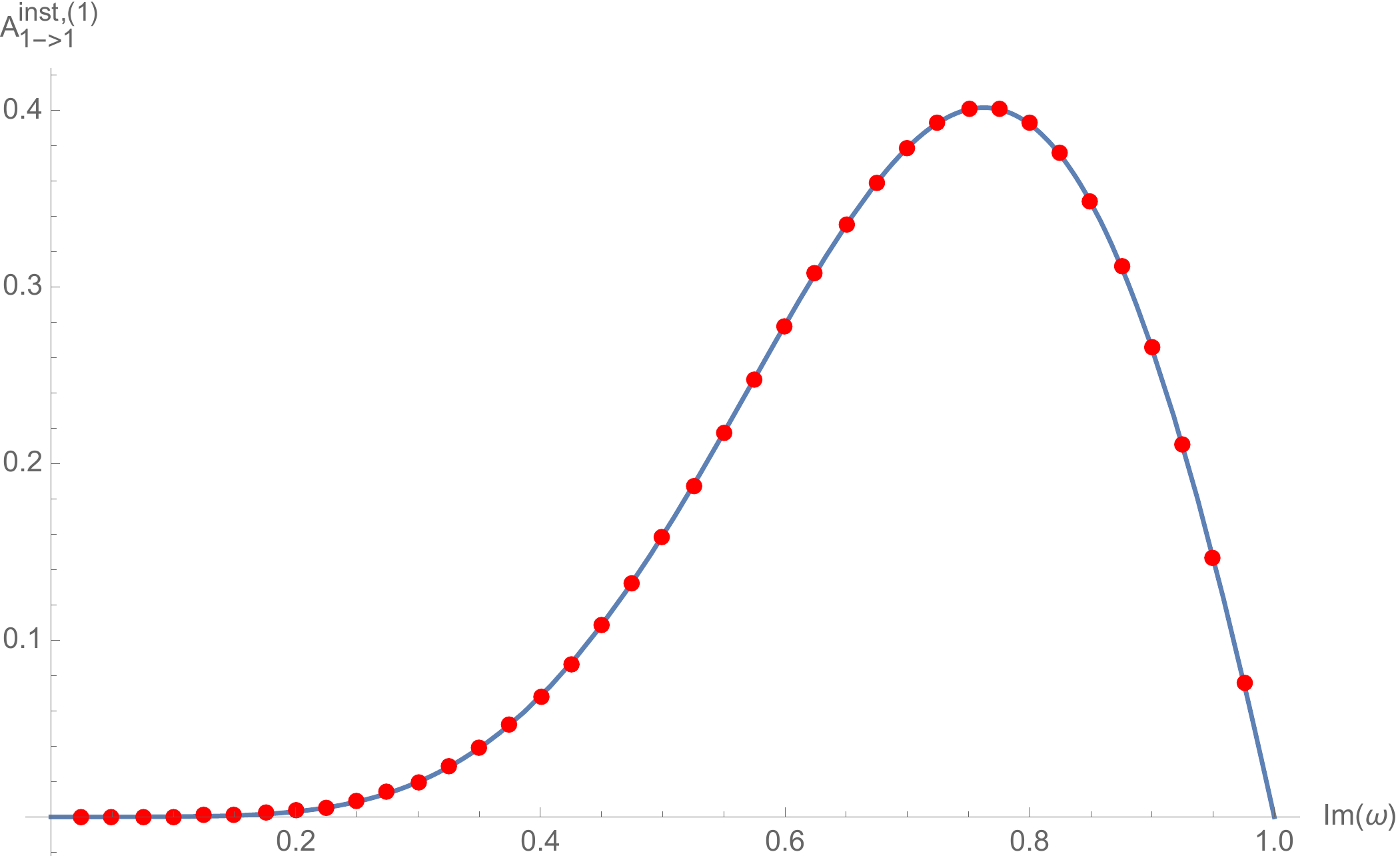}
\caption{In red dots is shown the non-perturbative correction of order $e^{-1/g_s} g_s$ to the $1\to 1$ closed string amplitude, computed from the worldsheet by numerically evaluating (\ref{eq:WSNLOall}) with the identification (\ref{scid}). In blue the matrix model result (\ref{eq:MMnlo}) for purely imaginary $\omega$ is plotted. 
}
\label{fig:WSvsMM}
\end{figure}

(\ref{eq:MMnlo}) can be compared to the worldsheet computation of disc 2-point amplitude and disc $\times$ annulus amplitude, which is given by the analytic continuation of the expression (\ref{eq:WSNLOall}) which is a priori valid for purely imaginary $\omega$ with $0<{\rm Im}(\omega)<1$. Note that for purely imaginary $\omega$, both (\ref{eq:WSNLOall}) and (\ref{eq:MMnlo}) are manifestly real. (\ref{eq:WSNLOall}) is evaluated numerically in this domain, up to the so-far-undetermined constants $S_{\rm ZZ}^{(1)}$ and $c'$. We find an excellent fit to (\ref{eq:MMnlo}) with\footnote{Here we have only indicated the error bar of the least squares fit, without taking into account the numerical error in the worldsheet computation itself.}
\ie\label{scid}
S_{\rm ZZ}^{(1)} = 0.496\pm0.001,~~~~~
c' = -1.399\pm 0.002.
\fe
The striking agreement between the worldsheet and matrix model results, provided the identification (\ref{scid}), is shown in Figure \ref{fig:WSvsMM}.


The agreement of the ZZ-instanton computation with the non-perturbative corrections in the matrix model amplitude distinguishes our non-perturbative completion of the matrix model dual from earlier proposals in the literature. For instance, as already mentioned, type 0B matrix model which amounts to filling both sides of the fermi sea up to $E\leq - \mu$ would lead to the reflection amplitude $(R(E))^*$ for the hole, rather than $(R(E))^{-1}$ \cite{Moore:1991zv, Takayanagi:2003sm, Douglas:2003up}, and would lead to different $\omega$-dependence in the order $e^{-2\pi \mu}$ amplitudes. Similarly, the theory of ``type I" considered in \cite{Moore:1991zv}, in which the reflection amplitude $R(E)$ is a phase, would also lead to different answers from the ZZ-instanton computation.\footnote{In fact, the theory of ``type I" has been argued in \cite{Polchinski:1994jp} to be non-perturbatively inconsistent with $c=1$ string theory, due to causality constraints on the scattering of ``tall" pulses.}

\section{Discussion}
\label{outlooksec}

In this paper we studied non-perturbative effects mediated by ZZ instantons in the worldsheet formulation of $c=1$ string theory, in which the Fischler-Susskind-Polchinski (FSP) mechanism plays an essential role in canceling open string divergences. We found that certain one-instanton amplitudes of closed strings at the leading and next-to-leading orders in $g_s$ are in agreement with the matrix model dual, provided that the closed string vacuum is identified with the state in the matrix quantum mechanics in which all scattering energy levels with no incoming flux from the other side of the potential are filled by free fermions/eigenvalues. Our proposal for the non-perturbative completion of $c=1$ string theory and its dual matrix model is explicitly different from that of type 0B string theory in two dimensions.

In our worldsheet computation of the $1\to 1$ closed string amplitude at order $e^{-1/g_s} g_s$, there are two unfixed constants. One of them, $S_{\rm ZZ}^{(1)}$, is formally computed by the worldsheet diagram of a 2-holed disc with no closed string insertions, which amounts to integrating a genus two Virasoro vacuum conformal block over a suitable moduli space. The latter is divergent near the boundary of the moduli space, due to the open string tachyon rather than the collective mode. It may be possible to regularize this divergence carefully so as to be compatible with the regularization of open string tachyon divergence of the annulus amplitude in section \ref{sec:anncomp}, and give a worldsheet derivation of the prediction (\ref{scid}) from the matrix model dual. The other constant $c'$ arises as a finite ambiguity in the cancelation of logarithmic divergences between diagrams of different topologies via the FSP mechanism. While $c'$ is eventually fixed by matching with the matrix model result, we do not know an a priori prescription for determining $c'$ in the worldsheet formalism.
\footnote{A similar ambiguity in the D-instanton effects in type IIB superstring theory was noted in \cite{Green:1997tv}.}

Note that while the perturbative expansion of the closed string amplitude in $c=1$ string theory is an asymptotic series, it is Borel summable, and the result of the Borel summation produces the ``perturbative part" of the amplitude captured by the function $K$ in (\ref{rrfac}). The ZZ-instanton effect gives rise to non-perturbative corrections on top of this. It will be interesting to see whether the multi-ZZ-instanton expansion reproduces the full non-perturbative amplitude of the matrix model, captured by the prefactor on the RHS of (\ref{rrfac}). It would also be desirable to recast the ZZ-instanton amplitudes in the language of open+closed string field theory \cite{Zwiebach:1997fe}, in such a way that the computation is manifestly finite, without the need for integration near the boundary of the moduli space. In doing so, one would presumably encounter finite ambiguities analogous to $c'$ in FSP cancelation mechanism at higher orders. A key question is how to fix these ambiguities based on the non-perturbative consistency of the string theory.\footnote{A possible clue lies in the matrix model result (\ref{ammoneton}): the non-perturbative correction to the fermion reflection phase $R(E)$ in (\ref{exactre}) is responsible for turning branch cuts on the complex $E$-plane into poles, as required by general properties of non-relativistic scattering, and cannot be modified arbitrarily (assuming that the dual description is a system of free fermions).}

The non-perturbative (exclusive) closed string amplitude, as we have seen, is not unitary. The unitary completion requires taking into account other asymptotic states, particularly those of fermions on the other side of the potential. The latter correspond to ZZ-branes in $c=1$ string theory with open string tachyon rolling to the ``wrong side" of the tachyon potential. It is conceivable that scattering amplitudes of closed string and ZZ-branes with rolling tachyon can be computed by worldsheet diagrams with ZZ-brane (rather than ZZ-instanton) boundary conditions, but a systematic formulation thereof is yet to be developed.

\section*{Acknowledgements}

We would like to thank Igor Klebanov, Juan Maldacena, Ashoke Sen, Marco Serone, Steve Shenker, Andy Strominger, and Yifan Wang for discussions. BB and VR thank the organizers of Bootstrap 2019 at Perimeter Institute, VR thanks the organizers of Strings 2019, Brussels, XY thanks the Galileo Galilei Institute for Theoretical Physics, Florence, the Center for Quantum Mathematics and Physics at University of California, Davis, International Center for Theoretical Physics, Trieste, and Nordic Institute for Theoretical Physics, Stockholm, for their hospitality during the course of this work. This work is supported in part by a Simons Investigator Award from the Simons Foundation, by the Simons Collaboration Grant on the Non-Perturbative Bootstrap, and by DOE grant DE-SC00007870. BB is supported by the Bolsa de Doutoramento FCT fellowship. VR is supported by the National Science Foundation Graduate Research Fellowship under Grant No. DGE1144152.

\appendix

\section{The $t\to0$ limit of the annulus 1-point diagram}
\label{sec:Cylt0}

In the $t\to 0$ limit, the annulus 1-point diagram has potential divergences due to propagation of on-shell closed string states that must be regularized. Changing the variable $t=1/s$ and using the modular property of $\eta$ and $\theta_1$, we can rewrite (\ref{eq:Cyl1}) as
\ie
\label{vomegas}
&\left\langle {\cal V}^+_{\omega} \right\rangle^{A^2}_{{\rm ZZ},x^0} 
\\
& = - 4\pi^2 g_s C_{A^2}e^{i\omega x^0}\int^\infty ds\, s^{-{3\over 2}} \eta(i s)\int_0^{\frac{1}{4}}dx \left[{2\pi e^{-4\pi s x^2 }\over is \partial_\nu\theta_1(0|is)}\theta_1(2i s x | is )\right]^{\frac{\omega^2}{2}}
\left\langle V_{P=\frac{\omega}{2}}(2\pi x)\right\rangle^{A^2(1/s)}_{\rm Liouviile,ZZ}.
\fe
In the $s\to \infty$ limit, we have ${2\pi \over \partial_\nu\theta_1(0|is)}\theta_1(2i s x | is ) \to 2 \sinh(2\pi x s)$,
and the Liouville annulus 1-point function reduces to an integral of disc 2-point functions (for $0<x<{1\over 4}$)
\ie{}
&\left\langle V_{P=\frac{\omega}{2}}(2\pi x) \right\rangle^{A^2(1/s)}_{\rm Liouviile,ZZ}
\\
&\to s^{2+\frac{\omega^2}{2}}\int_0^\infty\frac{dP_1}{\pi}\Psi^{ZZ}(P_1)e^{-2\pi s\left(P_1^2-\frac{1}{24}\right)}\frac{\left(2\tanh(\pi s x)\right)^{2+2P_1^2}}{\left(\sinh(2\pi s x)\right)^{2+\frac{\omega^2}{2}}}\left\langle V_{P=\frac{\omega}{2}}(i)V_{P_1}( y i)\right\rangle^{D^2}_{\rm Liouviile,ZZ},
\fe
where $y\equiv \tanh(\pi xs)$. Altogether, the large $s$ contribution to (\ref{vomegas}) can be written as
\ie
\left\langle {\cal V}^+_{\omega} \right\rangle^{A^2}_{{\rm ZZ},x^0}&\sim -4\pi^2 g_s C_{A^2}e^{i\omega x^0}2^{\frac{\omega^2}{2}}\int^\infty ds \, s^{1\over 2}\int_0^{\frac{1}{4}}dx { e^{-2\pi sx^2\omega^2}\over \sinh^2(2\pi x s)}\\
&~~~ \times\int_0^\infty\frac{dP_1}{\pi}\Psi^{ZZ}(P_1)e^{-2\pi s P_1^2} (2\tanh(\pi x s))^{2+2P_1^2}\left\langle V_{P=\frac{\omega}{2}}(i)V_{P_1}(i y)\right\rangle^{D^2}_{\rm Liouviile,ZZ}.
\label{eq:AmpT0}
\fe
The divergence comes from singular limits of the disc 2-point function on the RHS. The divergence in the small $y$ or small $xs$ limit was already analyzed in (\ref{eq:Cylx0}). There is also a potential divergence coming from the $y\to 1$ or large $xs$ limit, where 
\ie
\left\langle {\cal V}^+_{\omega} \right\rangle^{A^2}_{{\rm ZZ},x^0} \sim -4\pi^2 g_s C_{A^2}e^{i\omega x^0}\int^\infty ds s^\frac{1}{2}\int_0^\frac{1}{4}dx \int_0^\infty &\frac{dP_1dP_2}{\pi^2}\Psi^{{\rm ZZ}}(P_1)\Psi^{{\rm ZZ}}(P_2)C\left(\frac{\omega}{2},P_1,P_2\right)\\
&\times e^{-2\pi s\left[(1-2x)P_1^2+2x P_2^2-x\left(\frac{1}{2}-x\right)\omega^2\right] }.
\label{eq:Cyl1ptLargests}
\fe
In particular, if ${\rm Re}(\omega^2)>0$, the $s$-integral diverges for sufficiently small $P_1$ and $P_2$, and must be regularized in a way that preserves the expected analyticity in $\omega$. For numerical evaluation and comparison to the matrix model results, it is convenient to work with purely imaginary $\omega$, in which case the $s$-integral is manifestly convergent.

In the analytic continuation in $\omega$, one must ensure that the poles of the DOZZ structure constant, located at $\pm\frac{\omega}{2}+(P_1\pm P_2)=i n$ for nonzero integer $n$, 
do not cross the integration contour in $P_1$ and $P_2$, or equivalently, take into account of residue contributions when a pole crosses the integration contour. The simplest regime is $0<{\rm{Im}}(\omega)<1$, where such residue contributions are absent, and the regularized amplitude is given by (\ref{eq:Cyl1Imw}).

\section{Some numerical details}
\label{sec:detailnumerics}

\subsection{Disc 2-point diagram}

The disc 2-point Virasoro conformal blocks appearing in (\ref{bulkch}) and (\ref{bdrych}) are evaluated as an expansion in the elliptic nome $q=\exp[-\pi K(1-\eta)/K(\eta)]$, where $\eta$ is the cross-ratio of the correlator and $K(\eta)={}_2F_1(1/2,1/2,1;\eta)$, using Zamolodchikov's recurrence relations \cite{Zamolodchikov:1985ie}. Truncating the $q$-expansion to relatively low orders ($\sim 10-20$) is sufficient to obtain very accurate results. 

\begin{figure}[h!]
\centering
\includegraphics[width=0.7\textwidth]{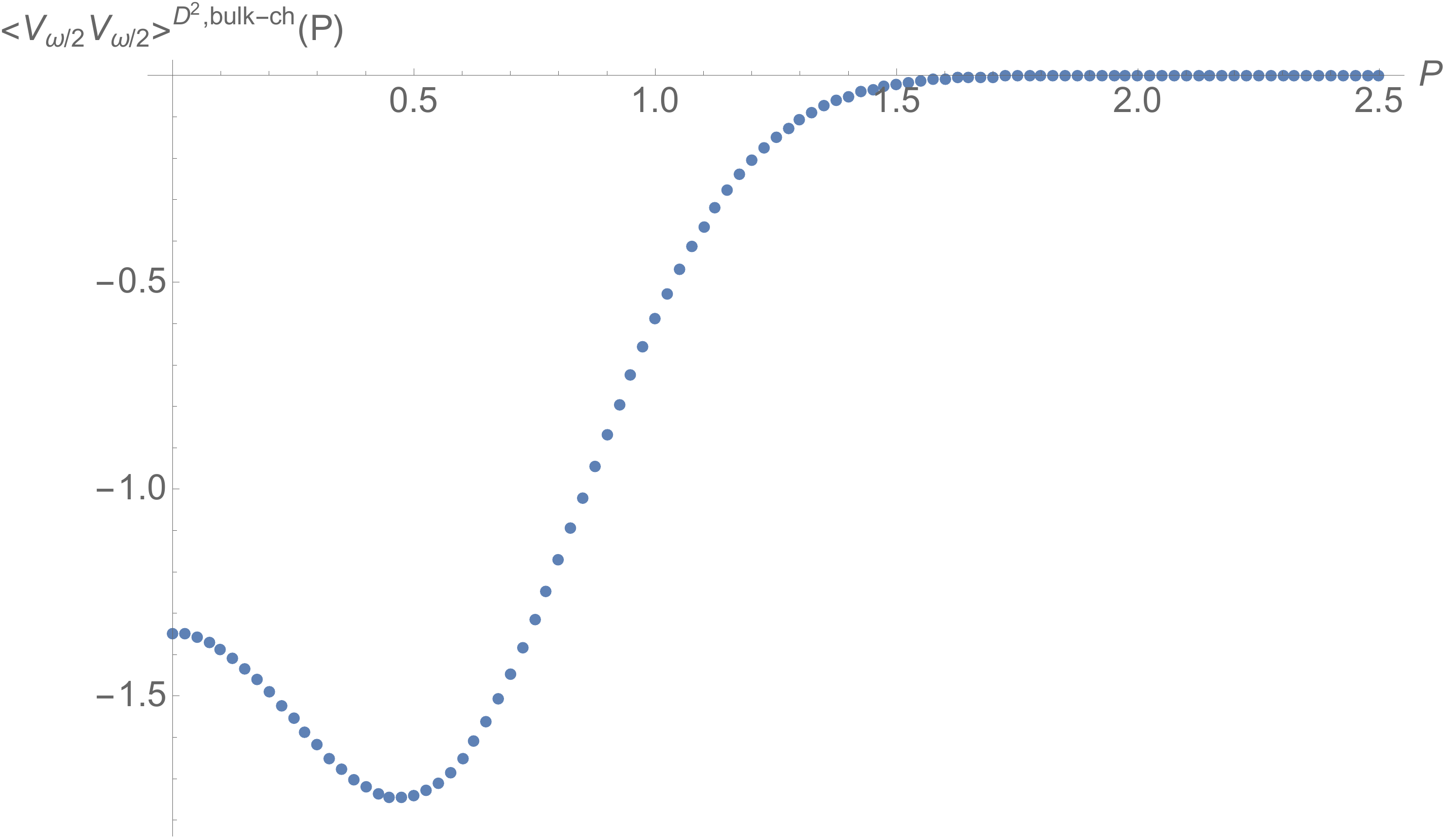}
\caption{A sample plot of the $P$-integrand in (\ref{eq:disk2reg}) from expanding the Liouville correlator in the bulk channel, after performing the $y$-integral for $1/2<y<1$. The energy of the incoming tachyon is taken to be $\omega=0.4i$.}
\label{fig:disk2ptBulkCh_sample1}
\end{figure}

\begin{figure}[h!]
\centering
\includegraphics[width=0.85\textwidth]{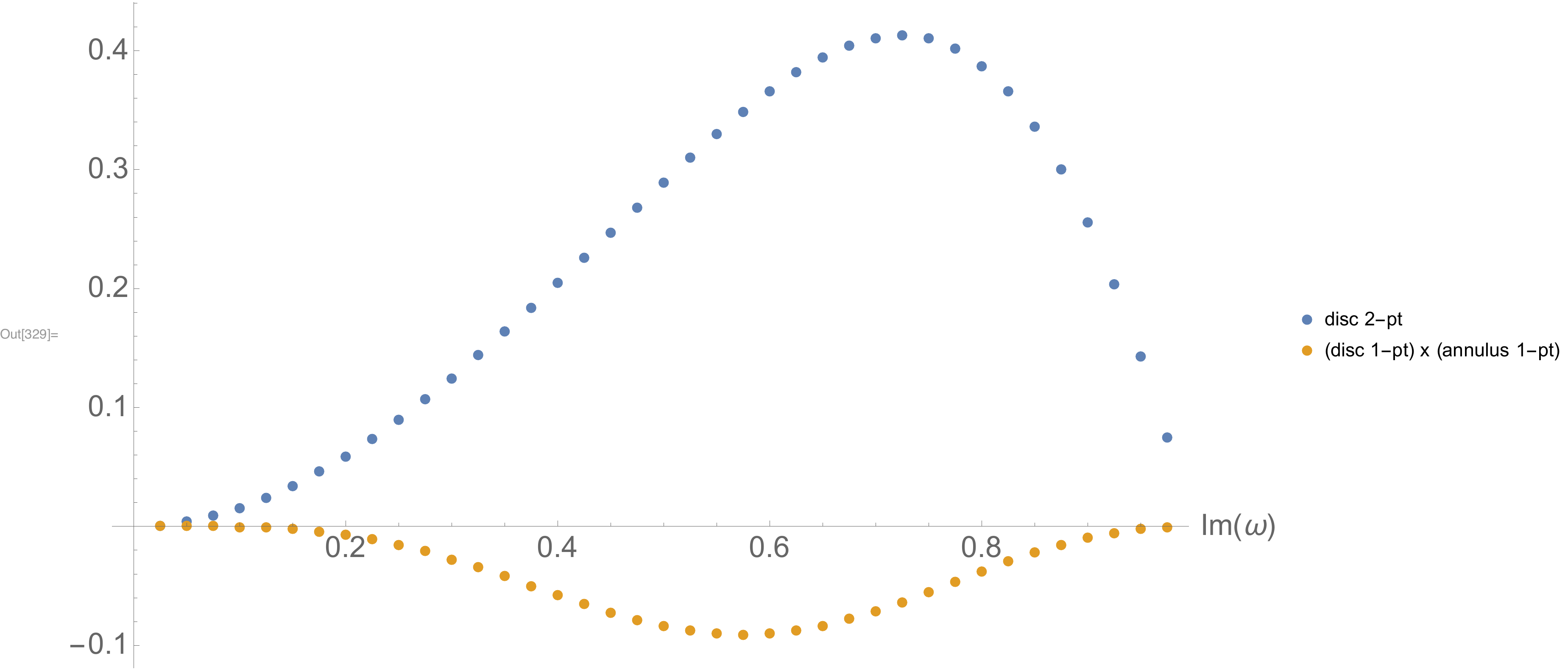}
\caption{Numerical results for the disc 2-point diagram (in blue) and the (disc 1-point)$\times$(annulus 1-point) diagram (in orange) contributions to (\ref{eq:WSNLOall}) (with the prefactor $e^{-1/g_s} g_s \delta(\omega-\omega')$ stripped off), at imaginary values of $\omega$ in the range $0<{\rm Im}(\omega)<1$.
}
\label{fig:DataDisk2ptAnnulus1pt}
\end{figure}

The numerical integration over the moduli space of the 2-punctured disc (\ref{eq:disk2reg}) is split into two regions: $0<y<{1\over 2}$, where the boundary channel conformal block representation is used, and ${1\over 2}<y<1$ where the bulk channel representation is used. In the boundary channel, only the vacuum channel block contributes, and the $y$-integral is straightforwardly evaluated after the subtraction in (\ref{eq:disk2reg}).
In the bulk channel, the conformal block decomposition involves an integration over the Liouville momentum $P$. It is convenient to first perform the $y$-integral and then the $P$-integral. In the vicinity of $y=1$, we can truncate the Virasoro conformal block to the lowest order in its $q$-expansion and perform the integration analytically as a function of $P$. The remaining of integration over $1/2< y<1$ is evaluated numerically at a sufficiently large set of sample values of $P$. We then numerically interpolate in $P$ and perform the $P$-integral. A sample plot of the $P$-integrand is shown in Figure \ref{fig:disk2ptBulkCh_sample1}. 
The regularized disc 2-point diagram (\ref{eq:disk2reg}), after integrating the collective coordinate, evaluated over a set of imaginary $\omega$ with $0<{\rm Im}(\omega)<1$, is shown in Figure \ref{fig:DataDisk2ptAnnulus1pt}.

\subsection{Annulus 1-point diagram}

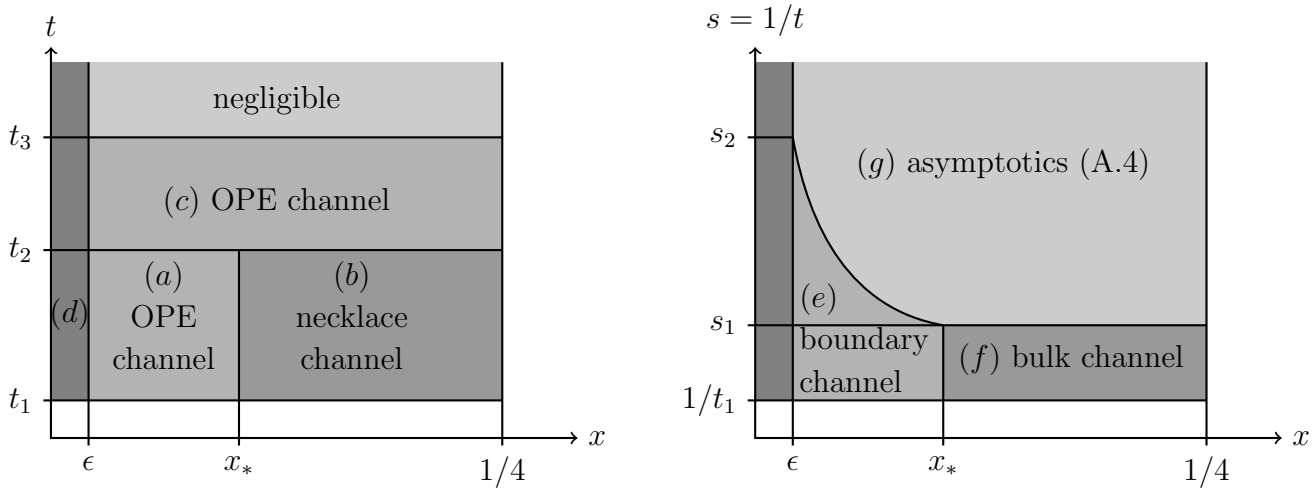
\begin{figure}[h!]
\centering
\subfloat{
\begin{tikzpicture}
\path [fill=black!50] (0,0.5) -- (0,5) -- (0.5,5) -- (0.5,0.5);
\path [fill=black!30] (0.5,0.5) -- (0.5,2.5) -- (2.5,2.5) -- (2.5,0.5);
\path [fill=black!30] (0.5,2.5) -- (0.5,4) -- (6,4) -- (6,2.5);
\path [fill=black!40] (2.5,0.5) -- (2.5,2.5) -- (6,2.5) -- (6,0.5);
\path [fill=black!20] (0.5,4) -- (0.5,5) -- (6,5) -- (6,4);
\draw [<->, thick] (0,5.2) -- (0,0) -- (7,0);
\draw (0,5.2) node[above] {$t$};
\draw (7,0) node[right] {$x$};
\draw [thick] (-0.1,0.5) -- (6,0.5);
\draw (-0.1,0.5) node[left] {$t_{1}$};
\draw [thick] (-0.1,2.5) -- (6,2.5);
\draw (-0.1,2.5) node[left] {$t_{2}$};
\draw [thick] (0.5,-0.1) -- (0.5,5);
\draw (0.5,-0.1) node[below] {$\epsilon$};
\draw [thick] (2.5,-0.1) -- (2.5,2.5);
\draw (2.5,-0.1) node[below] {$x_{*}$};
\draw [thick] (6,-0.1) -- (6,5);
\draw (6,-0.1) node[below] {$1/4$};
\draw [thick] (-0.1,4) -- (6,4);
\draw (-0.1,4) node[left] {$t_{3}$};
\node [align=center, below] at (1.5,2.5) {$(a)$\\ OPE\\ channel};
\node [align=center, below] at (4,2.5) {$(b)$\\ necklace\\ channel};
\node [align=center, below] at (3,3.5) {$(c)$ OPE channel};
\node [align=center] at (3,4.5) {negligible};
\node [align=center, below] at (0.25,2) {$(d)$};
\end{tikzpicture}
}~~~~
\subfloat{
\begin{tikzpicture}
\path [fill=black!50] (0,0.5) -- (0,5) -- (0.5,5) -- (0.5,0.5);
\path [fill=black!20] (0.5,1.5) -- (0.5,5) -- (6,5) -- (6,1.5);
\path [fill=black!40] (2.5,0.5) -- (2.5,1.5) -- (6,1.5) -- (6,0.5);
\path [fill=black!30] (0.5,0.5) -- (0.5,4) to [out=280,in=170] (2.5,1.5) -- (2.5,0.5);

\draw [thick] (0.5,4) to [out=280,in=170] (2.5,1.5);
\draw [<->, thick] (0,5.2) -- (0,0) -- (7,0);
\draw (0,5.2) node[above] {$s=1/t$};
\draw (7,0) node[right] {$x$};
\draw [thick] (-0.1,0.5) -- (6,0.5);
\draw (-0.1,0.5) node[left] {$1/t_{1}$};
\draw [thick] (-0.1,4) -- (0.5,4);
\draw (-0.1,4) node[left] {$s_{2}$};
\draw [thick] (0.5,-0.1) -- (0.5,5);
\draw (0.5,-0.1) node[below] {$\epsilon$};
\draw [thick] (2.5,-0.1) -- (2.5,1.5);
\draw (2.5,-0.1) node[below] {$x_{*}$};
\draw [thick] (6,-0.1) -- (6,5);
\draw (6,-0.1) node[below] {$1/4$};
\draw [thick] (-0.1,1.5) -- (6,1.5);
\draw (-0.1,1.5) node[left] {$s_1$};
\node [align=left, below] at (1.45,2.2) {$(e)$\\ boundary\\ channel};
\node [align=left, below] at (4.2,1.4) {$(f)$ bulk channel};
\node [align=center, below] at (3.3,4) {$(g)$ asymptotics (\ref{eq:Cyl1ptLargests})};
\end{tikzpicture}
}
\caption{Division of the moduli space of the 1-punctured annulus, for the purpose of numerical integration. The left figure covers the $t>t_1$ region whereas the right figure covers the $t<t_1$ region.}
\label{fig:AnnDomains}
\end{figure}

\begin{figure}[h!]
\centering
\begin{tabular}{c c c}
\includegraphics[width=0.4\textwidth]{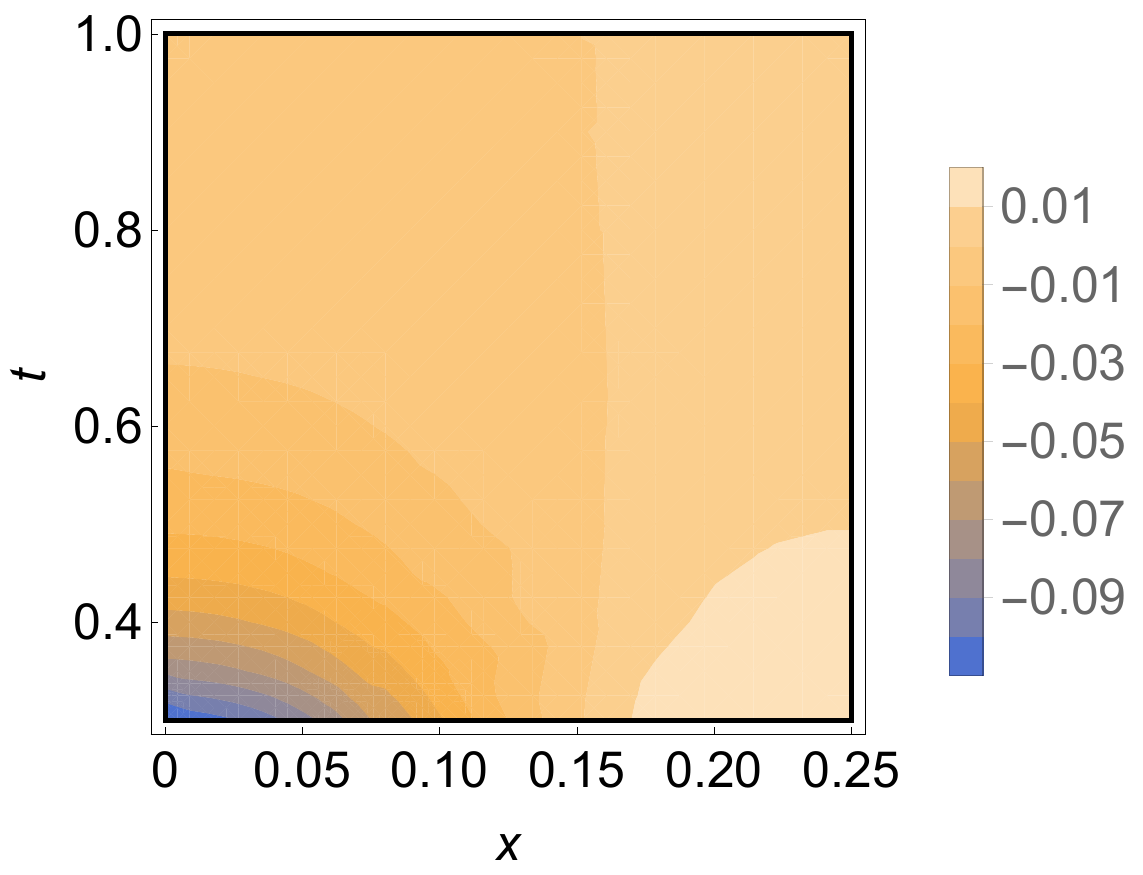}
& &
\includegraphics[width=0.42\textwidth]{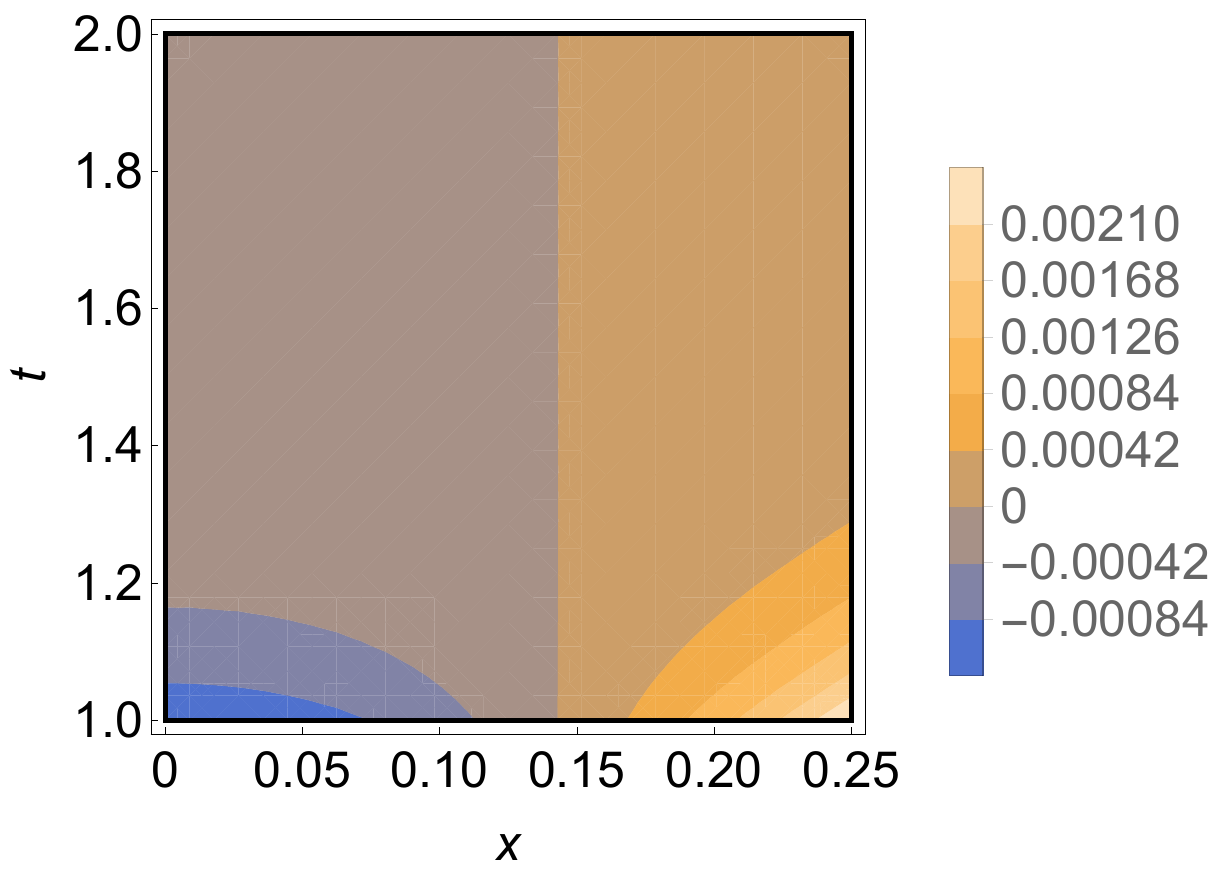} \\
regions (a) and (b)~~~~~ & & region (c)~~~~~~~~ \\
& & \\
\includegraphics[width=0.4\textwidth]{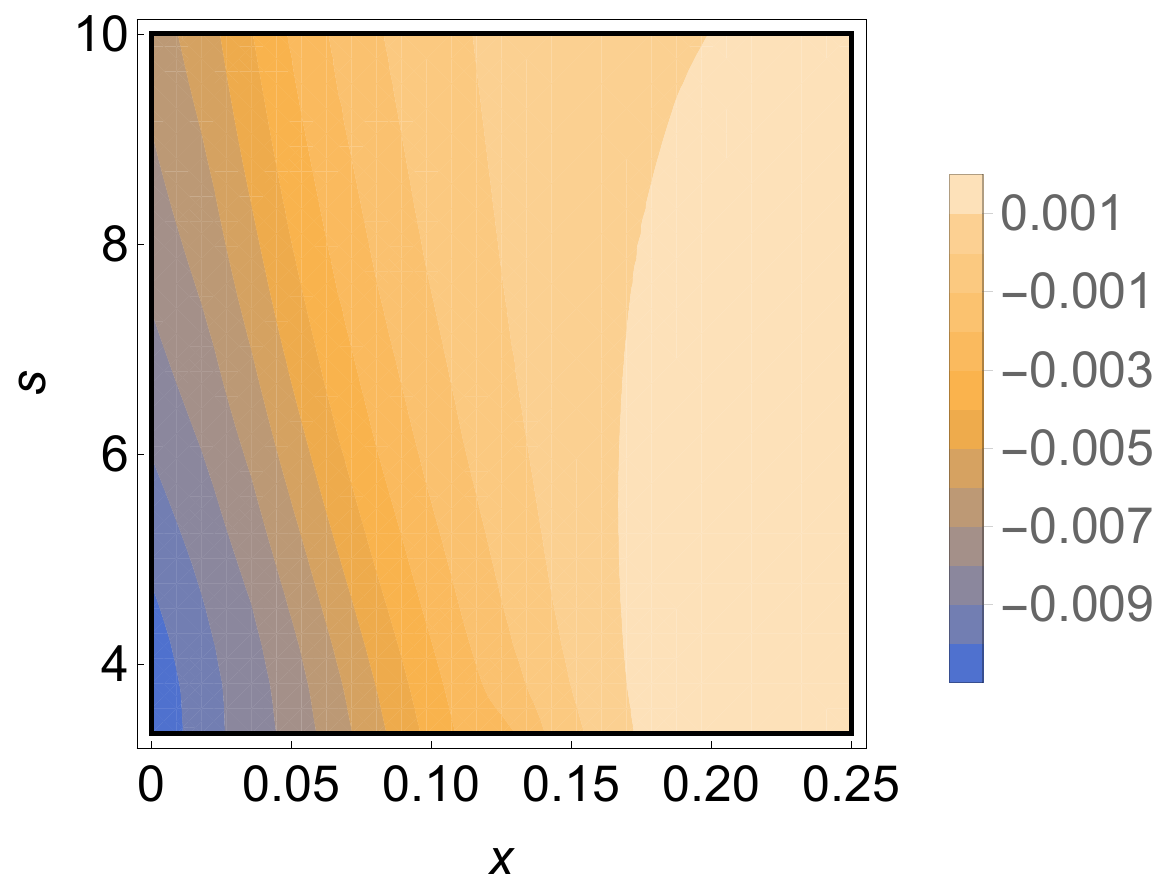}
& &
\includegraphics[width=0.42\textwidth]{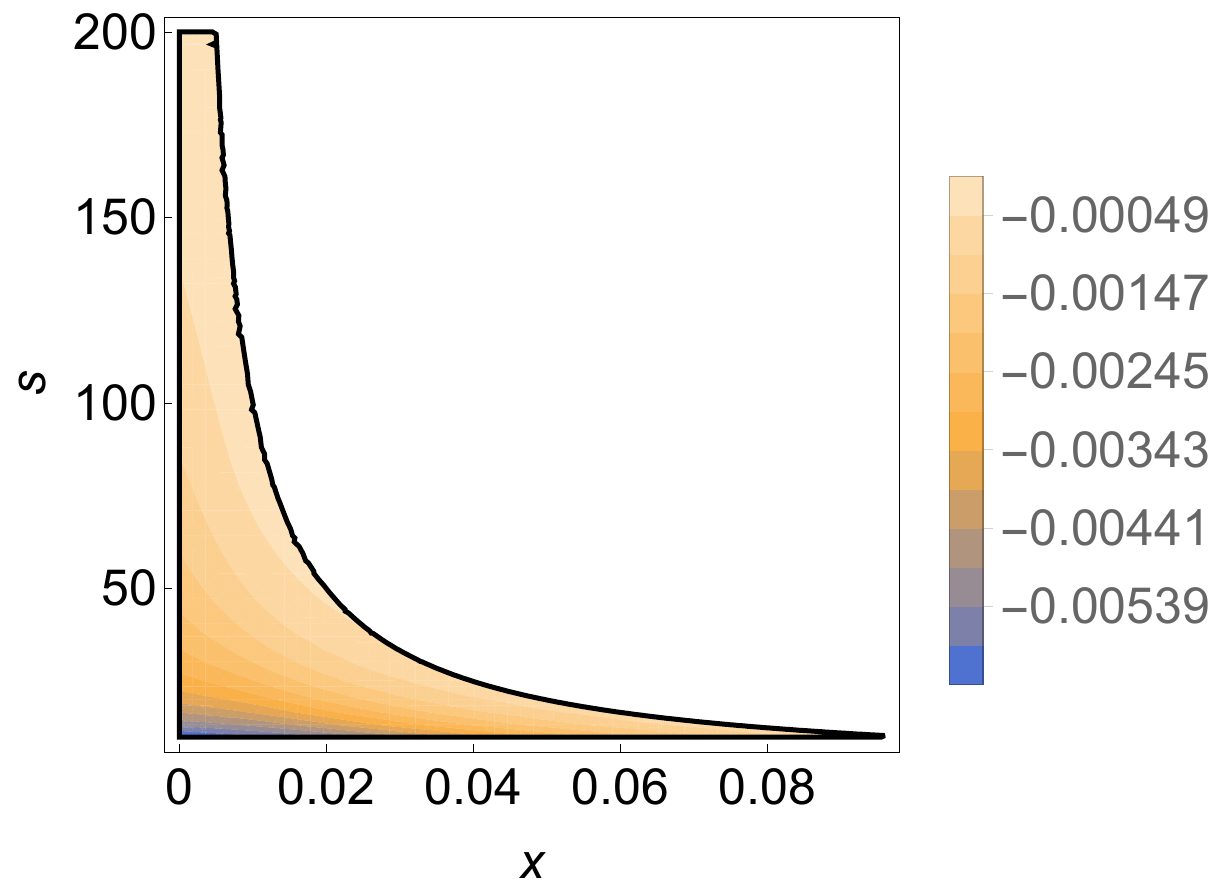} \\
region (f) and lower part of (e)~~~~~ & & upper part of (e)~~~~~
\end{tabular}
\begin{tabular}{c}
\includegraphics[width=0.5\textwidth]{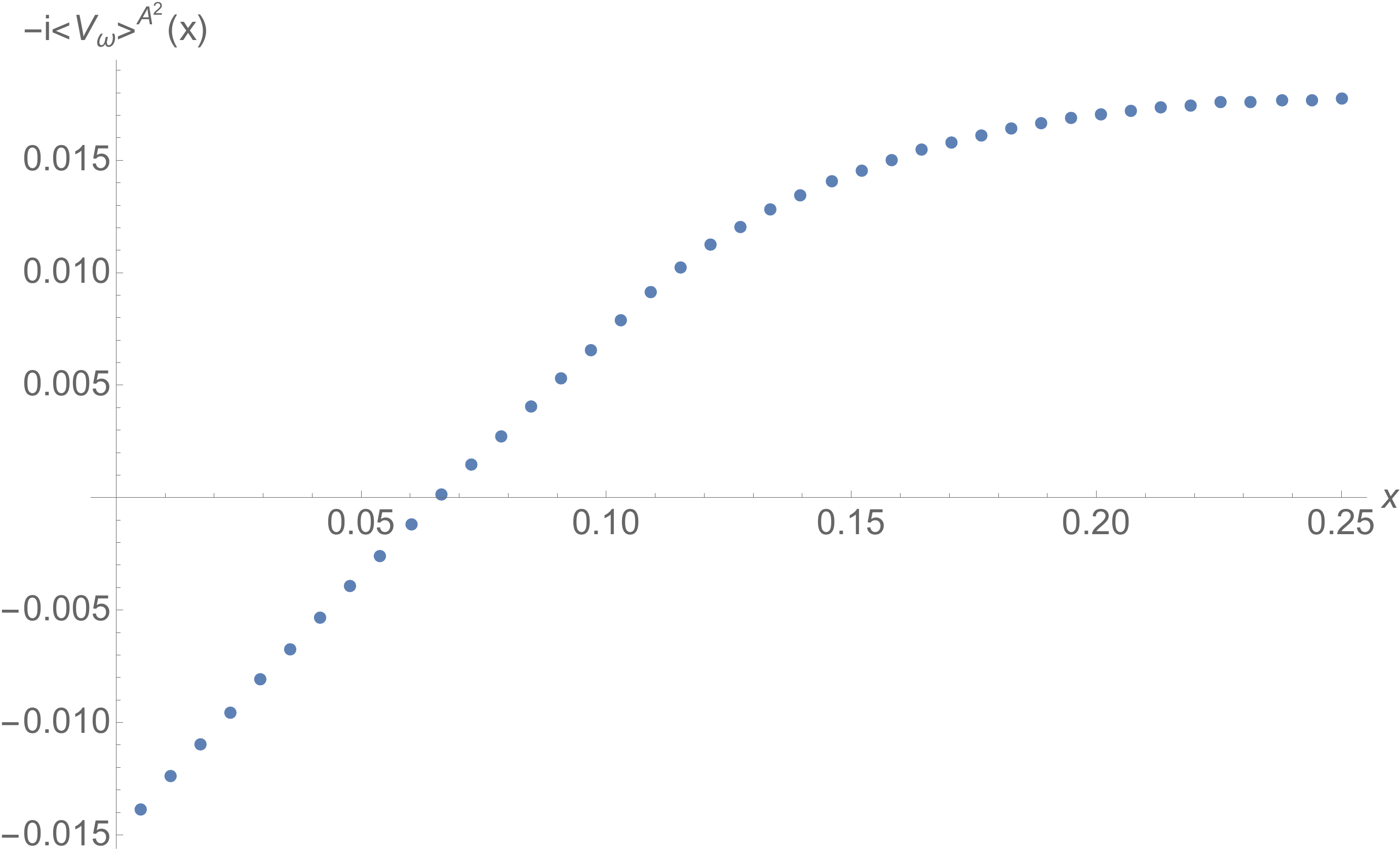}\\
region (g), after $s$-integration
\end{tabular}
\caption{Sample plots of the moduli space integrand for the annulus 1-point amplitude (\ref{eq:Cyl1Imw}) in various domains of Figure \ref{fig:AnnDomains}, evaluated at $\omega=0.75 i$. 
}
\label{fig:samplecontours}
\end{figure}

For the annulus 1-point diagram, we divide the moduli space into a number of domains. In each domain, a specific Virasoro conformal block decomposition and approximation scheme is used, as indicated in Figure \ref{fig:AnnDomains}. Sample plots of the integrand in various domains are shown in Figure \ref{fig:samplecontours}. 

For $t_1=0.3\leq t \leq t_2=1$, we express the Liouville annulus 1-point function in (a) the OPE channel (\ref{eq:CylOPE}) for $\epsilon=0.005\leq x\leq x_{*}=0.1$, and (b) the necklace channel (\ref{eq:Cylnecklace}) for $x_{*}<x\leq 1/4$. In each channel, the annulus 1-point conformal block, which is a special case of the torus 2-point block, is evaluated using the recurrence relations of \cite{Cho:2017oxl} and truncated in their expansion in the appropriate plumbing parameters (similarly to the evaluation of the genus one amplitude in \cite{Balthazar:2017mxh}). In the domain (c) $t_2\leq t \leq t_3 = 2$, it is convenient to use the OPE channel representation in which only the vacuum conformal block contributes. Finally, for sufficiently large $t$ the integrand becomes negligibly (exponentially) small and we can truncate the integration at $t=t_3$.

For smaller values of $t\leq t_1$ we can approximate the moduli integrand with the disc 2-point correlator as in (\ref{eq:AmpT0}). After passing to the variable $s=1/t$, we divide the $(x,s)$ integration domain according to the RHS of Figure \ref{fig:AnnDomains}. In region (f) with $x_{*}<x\leq 1/4$ and $1/t_1<s<s_1=10$, we expand the Liouville disc correlator in the bulk channel, whereas in region (e) with $1/t_1<s<1/x$ and $\epsilon\leq x\leq x_{*}$ we express the disc correlator in the boundary channel, using Zamolodchikov's recurrence relation and truncate in the $q$-series as before. In the region (g), one can further approximate the integrand using (\ref{eq:Cyl1ptLargests}) and perform the $s$-integral over $s> {\rm max} \{ s_1, 1/x \}$ analytically. 

Finally, in the slab region (d) $x<\epsilon$, the integrand becomes negligibly small whenever $t>t_2$ or $s>s_2$. For intermediate values of $t$ or $s$ we simply perform a linear extrapolation of the data points of adjacent regions, namely (a) and (e), down to $x=0$. 

Collecting contributions from all of these domains of the moduli space, the result for the regularized annulus 1-point diagram (\ref{eq:Cyl1Imw}), evaluated over a set of imaginary $\omega$, is shown in Figure \ref{fig:DataDisk2ptAnnulus1pt}.


\bibliographystyle{JHEP}
\bibliography{ZZinstdraft}

\end{document}